%% file: main.tex
\documentclass[acmsmall]{acmart}

\usepackage{float}
\usepackage{url}
\usepackage[normalem]{ulem}
\usepackage{graphicx}
\usepackage{balance}  
\usepackage{listings}
\usepackage{comment}
\usepackage{algpseudocode}
\usepackage[ruled,linesnumbered,vlined,noend]{algorithm2e}
\usepackage{amsmath}
\usepackage{graphics}
\usepackage{epsfig}
\usepackage{enumitem}
\usepackage{booktabs}
\usepackage{multirow}
\usepackage{ulem}
\usepackage{subfig}
\usepackage{lipsum}
\usepackage{array}
\usepackage{bbding}
\usepackage{algorithmicx}
\usepackage{algpseudocode}

\newtheorem{theorem}{Theorem}

\newcommand{\Paragraph} [1] {\smallskip\noindent{\bf #1. }}
\newcommand{\red}[1]{\textcolor{black}{#1}}
\newcommand{\blue}[1]{\textcolor{black}{#1}}

\newcommand{\oursys}{\texttt{GeoGauss}\xspace}
\newcommand{\kw}[1]{{\ensuremath {\mathsf{#1}}}\xspace}
\newcommand{\csn}{\kw{csn}}
\newcommand{\sen}{\kw{sen}}
\newcommand{\lsn}{\kw{lsn}}
\newcommand{\cen}{\kw{cen}}
\newcommand{\RR}{\kw{RR}}
\newcommand{\RC}{\kw{RC}}
\newcommand{\SI}{\kw{SI}}

\newcommand{\crdb}{\texttt{CRDB}\xspace}

\newcommand{\calvin}{\texttt{Calvin}\xspace}
\newcommand{\aria}{\texttt{Aria}\xspace}
\newcommand{\anna}{\texttt{Anna}\xspace}
\newcommand{\calvinfs}{\texttt{CalvinFS}\xspace}
\newcommand{\slog}{\texttt{SLOG}\xspace}
\newcommand{\qstore}{\texttt{Q-Store}\xspace}
\newcommand{\oursyss}{\texttt{GeoG-S}\xspace}
\newcommand{\oursysa}{\texttt{GeoG-A}\xspace}

\newcommand{\oursysha}{\texttt{GeoG-Raft}\xspace}
\newcommand{\oursyslb}{\texttt{GeoG-LB}\xspace}
\newcommand{\oursysrb}{\texttt{GeoG-RB}\xspace}
\newcommand{\calvinha}{\texttt{Calvin-Raft}\xspace}
\newcommand{\ariaha}{\texttt{Aria-Raft}\xspace}

\newcommand{\ie}{\emph{i.e.,}\xspace}
\newcommand{\eg}{\emph{e.g.,}\xspace}

\newcommand{\aka}{\emph{a.k.a.}\xspace}

\newcommand{\etc}{\emph{etc.}\xspace}

\AtBeginDocument{%
  \providecommand\BibTeX{{%
    \normalfont B\kern-0.5em{\scshape i\kern-0.25em b}\kern-0.8em\TeX}}}

\setcopyright{acmcopyright}
\setcopyright{acmlicensed}
\copyrightyear{2023}
\acmJournal{PACMMOD}
\acmYear{2023} \acmVolume{1} \acmNumber{1} \acmArticle{62} \acmMonth{5} \acmPrice{15.00}
\acmDOI{10.1145/3588916}

\begin{document}
\small
\normalsize


\title{GeoGauss: Strongly Consistent and Light-Coordinated OLTP for Geo-Replicated SQL Database}

%
\author{Weixing Zhou}
\orcid{0009-0000-9665-6052}
\author{Qi Peng}
\orcid{0009-0002-7952-6680}
\affiliation{
  \institution{Northeastern University}
  \streetaddress{No. 195, Chuangxin Road, Hunnan District}
  \city{Shenyang}
  \state{Liaoning}
  \country{China}
  \postcode{110169}
}
\email{zhouwx@stumail.neu.edu.cn}
\email{ffpq@stumail.neu.edu.cn}

\author{Zijie Zhang}
\orcid{0009-0006-3522-0910}
\affiliation{
  \institution{Huawei Technology Co., Ltd}
  \country{China}
}

\author{Yanfeng Zhang}
\orcid{0000-0002-9871-0304}
\authornote{Yanfeng Zhang is the corresponding author.}
\affiliation{
  \institution{Northeastern University}
  \streetaddress{No. 195, Chuangxin Road, Hunnan District}
  \city{Shenyang}
  \state{Liaoning}
  \country{China}
  \postcode{43017-6221}
}
\email{Zhangyf@mail.neu.edu.cn}

\author{Yang Ren}
\orcid{0009-0002-8679-8238}
\author{Sihao Li}
\orcid{0009-0007-4850-4258}
\affiliation{
  \institution{Huawei Technology Co., Ltd}
  \city{Xian}
  \state{Shanxi}
  \country{China}
}


\author{Guo Fu}
\orcid{0000-0001-7687-2815}
\author{Yulong Cui}
\orcid{0000-0001-8744-6117}
\affiliation{
  \institution{Northeastern University}
  \country{China}
}


\author{Qiang Li}
\orcid{0009-0009-8649-7655}
\affiliation{
  \institution{Huawei Technology Co., Ltd}
  \country{China}
}

\author{Caiyi Wu}
\orcid{0009-0009-9063-5723}
\author{Shangjun Han}
\orcid{0009-0008-9987-7625}
\author{Shengyi Wang}
\orcid{0009-0002-1157-4635}
\affiliation{
  \institution{Northeastern University}
  \country{China}
}


\author{Guoliang Li}
\orcid{0000-0002-1398-0621}
\affiliation{
  \institution{Tsinghua University}
  \streetaddress{30 Shuangqing Road}
  \city{Haidian District}
  \state{Beijing}
  \country{China}
}

\author{Ge Yu}
\orcid{0000-0002-3171-8889}
\affiliation{
  \institution{Northeastern University}
  \city{Shenyang}
  \state{Liaoning}
  \country{China}
}

\renewcommand{\shortauthors}{Weixing Zhou et al.}

\begin{abstract}
Multinational enterprises conduct global business that has a demand for geo-distributed transactional databases. Existing state-of-the-art databases adopt a sharded master-follower replication architecture. However, the single-master serving mode incurs massive cross-region writes from clients, and the sharded architecture requires multiple round-trip acknowledgments (\eg 2PC) to ensure atomicity for cross-shard transactions. These limitations drive us to seek yet another design choice. In this paper, we propose a strongly consistent OLTP database \oursys with full replica multi-master architecture. To efficiently merge the updates from different master nodes, we propose a multi-master OCC that unifies data replication and concurrent transaction processing. By leveraging an epoch-based delta state merge rule and the optimistic asynchronous execution, \oursys ensures strong consistency with light-coordinated protocol and allows more concurrency with weak isolation, which are sufficient to meet our needs. Our geo-distributed experimental results show that \oursys achieves 7.06X higher throughput and 17.41X lower latency than the state-of-the-art geo-distributed database CockroachDB on the TPC-C benchmark.
\end{abstract}



\received{July 2022}
\received[revised]{October 2022}
\received[accepted]{November 2022}

\maketitle

\input{intro}

\input{background}

\input{system}

\input{trans}

\input{impl}

\input{discuss}
\input{expr}

\input{related}

\begin{acks}
The work is supported by the National Natural Science Foundation of China (62072082, U2241212, 62202088, 62232009, 61925205, 62137001, 62272093), the National Social Science Foundation of China (21\&ZD124), CCF-Huawei Populus euphratica Innovation Research Funding (CCF-HuaweiDB2022003), the Fundamental Research Funds for the Central Universities (N2216012, N2216015), Huawei, TAL education, and Beijing National Research Center for Information Science and Technology (BNRist).
\end{acks}

\balance
\bibliographystyle{ACM-Reference-Format}
\bibliography{vldb_sample}

\end{document}

%% file: intro.tex
\section{Introduction}
\label{sec:intro}

Global companies have built their data centers located in many countries worldwide. To support their global business, it is desired to develop a geo-distributed transactional SQL database spread across multiple geographically distinct locations, \eg many telecom service providers have deployed their ICT databases under a geo-distributed setting.
The design goals are towards high availability, strong consistency, and high performance.

\begin{figure}[h]
\vspace{-0.2in}
  \centerline{\includegraphics[width=2.7in]{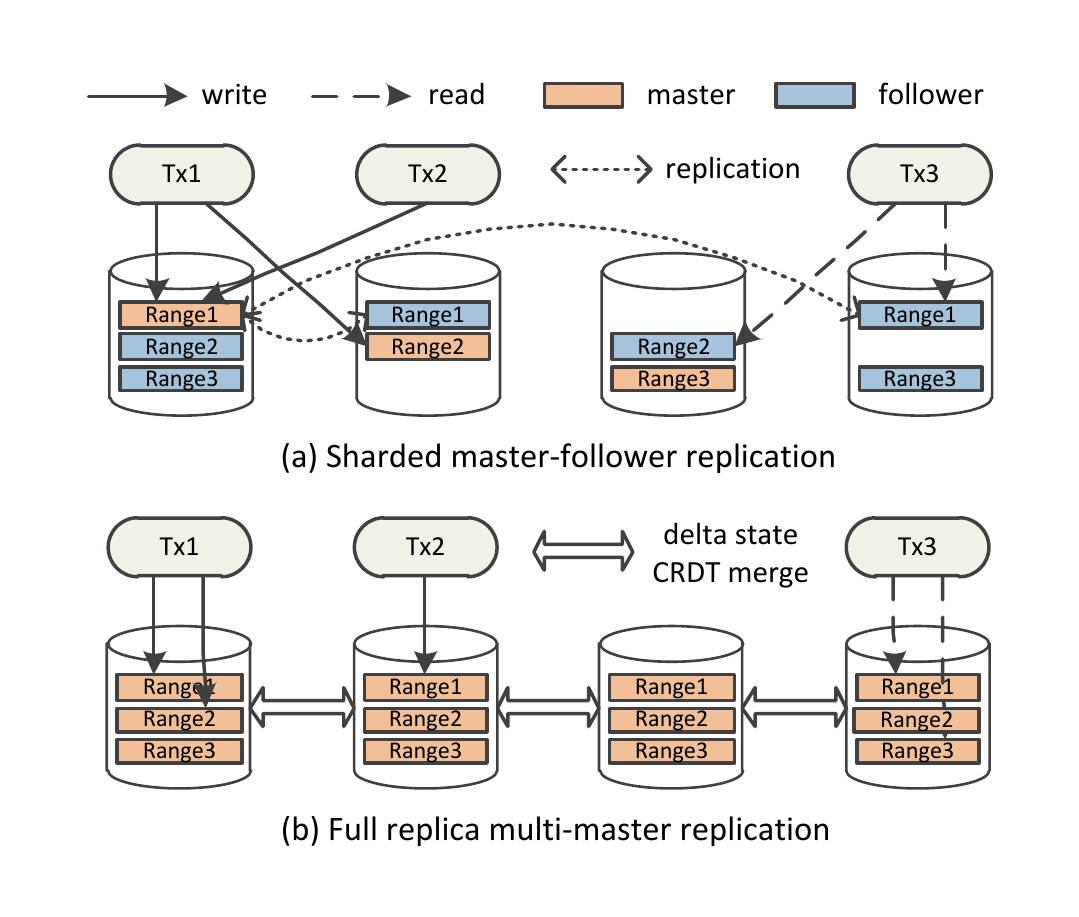}}
  \centering
  \vspace{-0.25in}
  \caption{Sharded master-follower replication vs. full replica multi-master replication.}
  \vspace{-0.1in}
  \label{fig:arch}
\end{figure}

High availability is usually achieved by redundant data replication, which is the process of storing the same data copies in multiple geographic zones. Data replication facilitates not only high availability but also geographic locality and read scalability, making data copies close to users at different regions to reduce read latency and to further improve overall data access throughput. Existing state-of-the-art geo-distributed transactional databases, 
\eg Google Spanner \cite{corbett2013spanner}, F1 \cite{10.14778/2536222.2536230}, CockroachDB \cite{taft2020cockroachdb}, YugabyteDB \cite{yugabytedb}, TiDB \cite{tidb}, \blue{\slog \cite{10.14778/3342263.3342647} and ConfluxDB \cite{chairunnanda2014confluxdb}} adopt a \textit{sharded master-follower replication} architecture as shown in Figure \ref{fig:arch}a. Data are partitioned into multiple shards according to the key range. Each shard is assigned to a single \textit{master} node serving all write/read requests, and it is replicated and placed to multiple geo-distributed \textit{follower} nodes serving only read requests. Due to its single-master architecture, write-write conflicts are gathered in the same worker to be easily coped with. In addition, sharding can disperse write requests to increase write throughput.

The sharded master-follower replication architecture is widely adopted \cite{corbett2013spanner,taft2020cockroachdb,tidb,yugabytedb,10.14778/3342263.3342647}, but it suffers from two major drawbacks. 1) The single-master serving mode requires to route the write requests from all clients to the single master node, which leads to cross-region writes and as a result increases transaction latency. Though this drawback can be alleviated by geo-aware partitioning and regional shard placement \cite{taft2020cockroachdb}, it still hurts performance, especially for applications without locality property. 
2) The sharded architecture relies on the two-phase commit (2PC) protocol to ensure atomicity. This requires multiple round-trip acknowledgments between the coordinator and the globally distributed workers, which further hurts performance.

Yet another choice for data replication is \textit{full replica multi-master} architecture as shown in Figure \ref{fig:arch}b, where each server maintains a full copy of data and all server nodes serve both read and write requests. By placing a full replica in each region, it can serve users with local writes/reads. With a full replica, 2PC is unnecessary to ensure atomicity. A number of multi-master systems emerge in recent years, \eg Aurora \cite{10.1145/3035918.3056101}, Riak \cite{riak}, Calvin \cite{thomson2012calvin}, 
FaunaDB \cite{fauna}, Anna \cite{wu2019autoscaling}, Aria \cite{lu2020aria} and Q-store \cite{qadah2020q}.
However, to employ multi-master architecture, there are three key challenges to be addressed.

\Paragraph{Challenge 1: Cross-Node Write-Write Conflicts} Different from the master-follower architecture where all writes to the same data are routed to the same node, with multi-master replication, concurrent updates to multiple replicas of the same data can result in cross-node write-write conflicts. Though this could be resolved with distributed two-phase locking protocol, it is too heavy in a geo-distributed setting. We address this challenge by employing \textit{Conflict-Free Replicated Datatypes} (CRDT) \cite{Shapiro:2011:CRD:2050613.2050642,wu2019autoscaling}. \blue{Specifically, we adopt an optimized state-based CRDT, \textit{delta-state CRDT} \cite{almeida2018delta, almeida2015efficient}, where only recently applied updates to a database state are disseminated instead of the entire database state. The state updates along with its local timestamp information are exchanged among all replicas. At the receiver side, the write-write conflicts are automatically merged by a function that joins any pair of replica updates and must be \textit{associative}, \textit{commutative}, and \textit{idempotent} (\ie \textit{ACI property}) \cite{Shapiro:2011:CRD:2050613.2050642}. The database state at any node is \textit{monotonically} increased by merging the state updates according to the same merge function, so that these replicas are updated independently without a global coordinator.} The ACI property of the merge function guarantees that the consistency of replicas can be eventually established after merging of all updates. However, eventual consistency is not acceptable in most transactional applications.

\Paragraph{Challenge 2: Strong Consistency} It is desirable to respond a committed/aborted status to the user as fast as possible, but 
this is not allowed before all replicas \red{reach} consistency. Linearizability, as one of the strongest single-object consistency (\ie all replicas \red{reach} the same state after every operation with real-time constraints), requires expensive coordination, while eventual consistency (\ie, all replicas \red{reach} the same state with no time constraint) which can be achieved by CRDT prompts performance without coordination but may incur various anomalies. \blue{We address this challenge by introducing \textit{epoch-based merge}, a compromise proposal (between real-time synchronous merge and asynchronous merge) that guarantees consistency at the granularity of \textit{epochs}.}
\blue{Unlike 2PC, where each transaction has a coordinator, we only coordinate at the granularity of epochs, so the coordination is amortized for a set of transactions.} 
After collecting the updates of an epoch from all nodes, the identical set of updates are merged on each node according to the same function with ACI property, and then applied to the previously consistent local replica state. Thus, a globally consistent snapshot can be reached among all nodes on a per-epoch basis. 

\Paragraph{Challenge 3: Performance} 
However, by employing the epoch-based merge, the synchronization barriers that require to receive updates from all peers are introduced (the only coordination), which hurts performance a lot. \blue{To address this limitation, we employ optimistic execution under a multi-master setting. The master nodes \textit{optimistically} execute their local SQL requests of epoch $i$ based on the current database state, which is not necessarily the most recent epoch snapshot ($i-1$).} The resulting write sets (\aka updates) are exchanged among nodes for epoch-based conflict merge. \blue{The merge results are used for generating a new snapshot $i$.} 
Only those successfully validated transactions are allowed for commitment. In other words, the transaction execution phase is performed \textit{asynchronously} among nodes regardless of epoch concept, while the validation phase has to be performed \textit{synchronously} among nodes based on the most recent snapshot. On the other hand, considering that the mainstream sharded master-follower systems \cite{corbett2013spanner,taft2020cockroachdb,yugabytedb} support \textit{serializability} (which executes a transaction’s logic as a single unit), \blue{the optimistic execution allows for overlapping the read and write sets across transactions with weak isolation and as a result brings more concurrency and performance.} This greatly meets the requirements of ICT databases for telecom providers on strong replica consistency and high throughput where weak isolation is sufficient in most scenarios.

By integrating the above techniques, we propose \textit{multi-master OCC}, an epoch-based optimistic transaction processing scheme under \red{a} multi-master setting. It unifies data replication with optimistic concurrency control, supporting multiple isolation levels with high concurrency while guaranteeing strong replica consistency at the granularity of epochs. Furthermore, we build a multi-master geo-replicated OLTP database \oursys by modifying openGauss MOT \cite{mot}, an in-memory storage engine that is highly optimized for multi-core processors. \oursys supports a full-featured SQL engine with delta-state CRDT merge. To the best of our knowledge, \oursys is the first to integrate CRDT technique into commercial SQL databases with full SQL \red{support}. 

To sum up, we make the following three key contributions.
\begin{itemize}[leftmargin=*]
    \item \textbf{Epoch-based Multi-Master OCC.} By relying on delta-state CRDT and epoch-based replication, we propose a multi-master OCC protocol that supports \blue{light-coordinated} high throughput transactions with strong consistency and supports multiple isolation levels.
    \item \textbf{\oursys Database.} We develop a geo-replicated OLTP database \oursys with full SQL \red{support} and rich optimizations (\eg high concurrency, efficient communication, pipelining, and fault tolerance) to adapt to a geo-distributed environment.
    \item \textbf{Extensive Experiments.} We conduct extensive experiments under a geo-distributed environment with YCSB and TPC-C benchmarks. We compare {\oursys} with \blue{CockroachDB (CRDB) \cite{taft2020cockroachdb}, Calvin \cite{thomson2012calvin}, Aria \cite{lu2020aria}, CalvinFS \cite{thomson2015calvinfs}, Q-Store \cite{qadah2020q}, SLOG \cite{10.14778/3342263.3342647}, and a coordination-free KV database Anna \cite{wu2019anna}.}
    Our results show that \oursys achieves 1.11X-7.06X higher throughput and 2.28X-17.41X lower latency than these competitor systems on the TPC-C benchmark and 0.44X-87.36X higher throughput and 0.27X-30.94X lower latency on a medium-contention YCSB benchmark.
\end{itemize}

%% file: background.tex
\section{Background}
\label{sec:2}

\begin{table*}[t]
	\vspace{-0.1in}
	\caption{Summary of replicated systems}
	\vspace{-0.1in}
	\label{tab:system}
	\centering
	\small
	{
        \resizebox{\textwidth}{29mm}{\begin{tabular}{cccccccc}
		 \toprule
		{\textbf{System}} &
		{\textbf{Model}} &
		{\textbf{Replication}} &
		{\textbf{Rep. Unit}} &
		{\textbf{Ordering}} &
		{\textbf{Consistency}}&
		{\textbf{CC}}&
		{\textbf{Isolation}} \\
		\hline
		{\textbf{openGauss} \cite{opengauss}} & SQL & master-follower & redo log & TSO & semi-sync. (eventual) & MV2PL & SI \\
		{\textbf{openGauss MOT} \cite{mot}} & SQL & master-follower & redo log & TSO & semi-sync. (eventual) & OCC & RR \\
		{\textbf{TiDB} \cite{tidb}} & SQL & master-follower & binary log  & TSO & quorum (linearizability) & MV2PL  & SER \\
		{\textbf{Spanner} \cite{corbett2013spanner}} & SQL & master-follower & redo log & TrueTime & quorum (linearizability) & MV2PL  & SER \\
		{\textbf{CockroachDB} \cite{taft2020cockroachdb}} & SQL & master-follower & binary log  & HLC & quorum (linearizability) & MVTO & SER \\
		{\textbf{HBase} \cite{hbase}} & KV & master-follower & HFile & \blue{local time} & semi-sync. (eventual)& MV2PL  & RC \\
		{\textbf{DynamoDB} \cite{decandia2007dynamo}} & KV & masterless & KV & \blue{vector clock} & quorum (sequential) & last write wins  & \blue{None} \\
		{\textbf{Cassandra} \cite{lakshman2010cassandra}} & KV & masterless & KV & \blue{local time} & quorum (sequential) & last write wins  & \blue{None} \\
		{\textbf{Bitcoin} \cite{nakamoto2008bitcoin}} & ledger & multi-master & block of txs & PoW & quorum (sequential) & serial  & \blue{SER}\\
		{\textbf{Fabric} \cite{androulaki2018hyperledger}} & ledger & multi-master & block of txs & ordering service & quorum (sequential) & MVOCC & \blue{SER}\\
		{\textbf{Anna} \cite{wu2019anna}} & KV & multi-master & KV & commutative & CALM/CRDT (eventual) & conflict free
		 & RC \\
		 {\textbf{Redis CRDT} \cite{redis}} & KV & multi-master & KV & commutative & CRDT (eventual) & conflict free
		 & \blue{None} \\
		{\textbf{Calvin} \cite{thomson2012calvin}} & SQL* & multi-master & batch of SQLs & local time & deterministic (sequential) & ordered locks  & SER \\
		{\textbf{Aria} \cite{lu2020aria}} & SQL* & multi-master & batch of SQLs & local time & deterministic (sequential) & dep. graph  & SER \\
		{\textbf{\oursys} (ours)} & SQL & multi-master & batch of write sets & local time & epoch CRDT (sequential) & multi-master OCC  & SI \\ 
		\bottomrule
	\end{tabular}
	}
	}
	\vspace{-0.2in}
	\normalsize
\end{table*}

Geo-replicated systems exhibit three key advantages, high availability, geographic locality, and read scalability. This section reviews existing replicated systems from different dimensions.

\subsection{Replicated Data Systems}
\label{sec:2:repsys}
Table \ref{tab:system} summarizes multiple replicated systems with their key features. We next study the key techniques used for data replication.

\subsubsection{Replication Architectures} There are three main replication architectures, \textit{master-follower}, \textit{multi-master}, and quorum-based \textit{masterless} architecture. A number of production systems (\eg openGauss \cite{opengauss}) employ master-follower replication mainly for redundant backup support, while another set of systems employ master-follower replication for scaling read throughput (\eg CockroachDB \cite{taft2020cockroachdb}, \blue{ConfluxDB \cite{chairunnanda2014confluxdb}}) by allowing follower replica to serve read requests.
However, it is required to route all users' write requests to the single master server, which is unsuited for geo-distributed \red{databases} where users are spread across regions. 
With masterless (or leaderless) replication, a user's write/read operation is sent to all replicas, and a quorum protocol is used to avoid stale read by comparing the monotonic version number (\eg DynamoDB \cite{decandia2007dynamo}, Cassandra \cite{lakshman2010cassandra}). However, all users' write requests are required to be sent to multiple geo-remote servers, which is costly in geo-distributed scenarios. With multi-master replication \cite{couchdb,arangodb,cloudant,10.1145/1217935.1217947,10.14778/3342263.3342270,galera,extremedb}, all nodes can serve both read and write requests for their local users. It is naturally adapted to geo-distributed requirements.

\blue{In addition, supporting distributed transaction processing on Byzantine Fault Tolerant (BFT) and Crash Fault Tolerant (CFT) clusters has been studied in blockchain systems. For example, ResilientDB \cite{gupta2020resilientdb} proposes a hierarchical multi-master architecture for geo-scale deployments. And there also exist many sharding blockchains which provide scalability by grouping nodes into clusters and partitioning data among several independently-run clusters, \eg AHL \cite{dang2019towards}, Caper \cite{amiri2019caper}, SharPer \cite{amiri2021sharper}, RingBFT \cite{rahnama2021ringbft} and ByShard \cite{hellings2021byshard}. Most of these works focus on optimizing the performance of cross-shard transactions which is also a bottleneck in distributed DBMSs \cite{harding2017evaluation}. As an alternative, dynamically transferring the mastership of data to avoid expensive multi-shard coordination can improve the performance, \eg DynaMast \cite{abebe2020dynamast} and MorphoSys \cite{abebe2020morphosys}.}

\subsubsection{Ordering and Consistency of Replicas}
With master-follower architecture, some databases use replication only for backup (\eg MySQL \cite{mysql}, HBase \cite{hbase}, and openGauss \cite{opengauss}).
\blue{These systems use different ordering techniques due to different requirements, \eg openGauss relies on a centralized \textit{timestamp oracle (TSO)} service to achieve snapshot isolation, HBase uses local time for identifying the version of a value.}
To maximize performance, they are configured with asynchronous replication or semi-synchronous replication \cite{semisync} (using synchronous replication for a subset of the followers). This means that the consistency of all replicas \red{may be} not guaranteed at a certain time point (eventual consistency). 
Another set of sharded databases with master-follower replication provide serializability and linearizability.
For example, Google Spanner \cite{corbett2013spanner} relies on \textit{TrueTime} 
, CockroachDB \cite{taft2020cockroachdb} relies on \textit{hybrid logical clock (HLC)}, and OceanBase \cite{oceanbase} and TiDB \cite{tidb} rely on a centralized \textit{timestamp oracle (TSO)} service. The replica consistency is guaranteed with leader-based protocols, \eg Paxos and Raft. \blue{ISS \cite{stathakopoulou2022state} was recently proposed to improve these leader-based consensus protocols by efficiently multiplexing consensus instances for scalability.}

Many NoSQL KV databases adopt masterless replication without stable master and use quorum protocol to ensure consistency, \blue{ such as DynamoDB \cite{decandia2007dynamo} detects updated conflicts by vector clock and Cassandra \cite{lakshman2010cassandra} uses the local time to identify the version of a value.}
However, this \textit{quorum-based} approach without a stable master pays for it in performance. Hence, masterless databases are developed either for use cases that can tolerate eventual consistency or pays for strong consistency in performance,  \eg DynamoDB \cite{decandia2007dynamo}, Riak \cite{riak}, and Cassandra \cite{lakshman2010cassandra}.

For multi-master architecture, MySQL and PostgreSQL both provide tools for cross-site multi-master replication, \eg MySQL Tungsten \cite{tungsten} and PostgreSQL BDR \cite{pgbdr}, but they use asynchronous replication and fail to guarantee strong consistency. The blockchain system is a particular multi-master replication system with Byzantine-fault tolerance. The key to ensuring consistency is to reach a consensus on the order of transactions under a trustless environment. 
For example, permissionless blockchain Bitcoin \cite{nakamoto2008bitcoin} employs \textit{Proof-of-Work} (PoW), permissioned blockchain Fabric \cite{androulaki2018hyperledger} leverages an ordering service to determine a global order of transactions.
In this way, these blockchains can achieve sequential consistency \blue{and serializable isolation}.

Different from others, coordination-free replication, \eg Berkeley Anna \cite{wu2019anna} and Redis CRDT \cite{redis}, utilizes mathematical properties of operations (\ie \textit{associative}, \textit{commutative}, and \textit{idempotent}) to reach replica consistency with the out-of-order transactions as input on different replicas. Typical works include \textit{Consistency As Logical Monotonicity (CALM)} \cite{bailis2015coordination, alvaro2011consistency, 10.14778/2735508.2735509,  Conway:2012:LLD:2391229.2391230, 10.1145/3110214} and \textit{Conflict-free Replicated Data Type (CRDT)} \cite{shapiro2011conflict, preguicca2018conflict, almeida2018delta}. 
Another consistency guarantee approach is adopted by \textit{deterministic databases}. Multiple master nodes accept transaction requests independently and exchange them between each other with batches to achieve replication. On each replica, the identical received set of transactions are executed in batches in a deterministic order. Before execution, all nodes have made \red{an} agreement on the ordering rule, \eg according to the globally unique ID (local time + server ID). Essentially, it guarantees sequential consistency and supports serializable isolation, \eg Calvin \cite{thomson2012calvin}, Aria \cite{lu2020aria}, \blue{\calvinfs \cite{thomson2015calvinfs} and \qstore \cite{qadah2020q}}.

\subsubsection{Replication with Transaction Processing} 
Traditional concurrency control techniques, such as \textit{multi-version two-phase locking} (MV2PL) and \textit{multi-version timestamp ordering} (MVTO), can be directly applied in single master architecture since all concurrent writes are processed locally. In sharded databases, 2PC is required for cross-shard transactions to ensure atomicity and consistency. For KV stores that base on masterless replication, they follow the last-write-win rule to resolve write-write conflicts.

For multi-master replication, there exist various conflict resolution approaches for concurrent updates. The most special one is Anna \cite{wu2019anna} which is naturally conflict-free since the update operations are required to be insensitive to the execution order, \eg set union and counter. In Bitcoin \cite{nakamoto2008bitcoin}, the single node that first solves a PoW puzzle executes transactions serially without concurrency. In Fabric \cite{androulaki2018hyperledger}, all peer nodes use \textit{multi-version optimistic concurrency control (MVOCC)} to execute their local transactions parallelly, then based on the order consensus they validate their executions and abort conflict transactions. 
Deterministic databases execute transactions according to a predefined serial order, which \blue{have} limitation on concurrency since the operating system schedules concurrently running threads in a fundamentally non-deterministic way. Existing deterministic databases rely on dependency graphs (Aria \cite{lu2020aria}) or ordered locks (Calvin \cite{thomson2012calvin}) to provide more parallelism while ensuring determinism. 

\subsection{Requirements}
\label{sec:2:requirement}

\blue{Despite many distributed databases having been proposed, the demand of top telecom customers drives us to propose a new OLTP database that fulfills the following requirements.}

\begin{itemize}[leftmargin=*]
    \item \textbf{\blue{Multi-Master Architecture.}} \blue{As motivated in Section \ref{sec:intro}, multi-master architecture is more suited for a geo-distributed setting, which offers low latency service.}
    \item \textbf{\blue{Full-Featured SQL Engine.}} \blue{High SQL coverage and interoperability are essential for a commercial database and can meet the needs of various users, \eg OLTP, OLAP, and stored procedures.}
    \item \textbf{\blue{Strong Consistency.}} \blue{As a geo-replicated data system, strong consistency of replicas is desired. However, \textit{linearizability} is too expensive and unnecessary. A little real-time property can be compensated for performance, because \textit{sequential consistency} is sufficient in most scenarios, \eg online retail and global trading.}
    \item \textbf{High Performance with Weak Isolation.} Many ICT databases used by telecom service providers are deployed under geo-replicated settings, such as Operation Support Systems (OSS), Customer Relationship Management (CRM), and Enterprise Resource Planning (ERP). They demand high throughput and low latency but do not require strong isolation. It is sufficient to support weaker isolation rather than serializability.

\end{itemize}

Existing geo-distributed transactional databases pay too much for linearizability and serializability, which cannot satisfy our requirements. Next, we will present \oursys.

%% file: system.tex
\vspace{-0.2in}
\section{System Overview}
\label{sec:3}

\begin{figure}[h]
\vspace{-0.3in}
  \centerline{\includegraphics[width=3in]{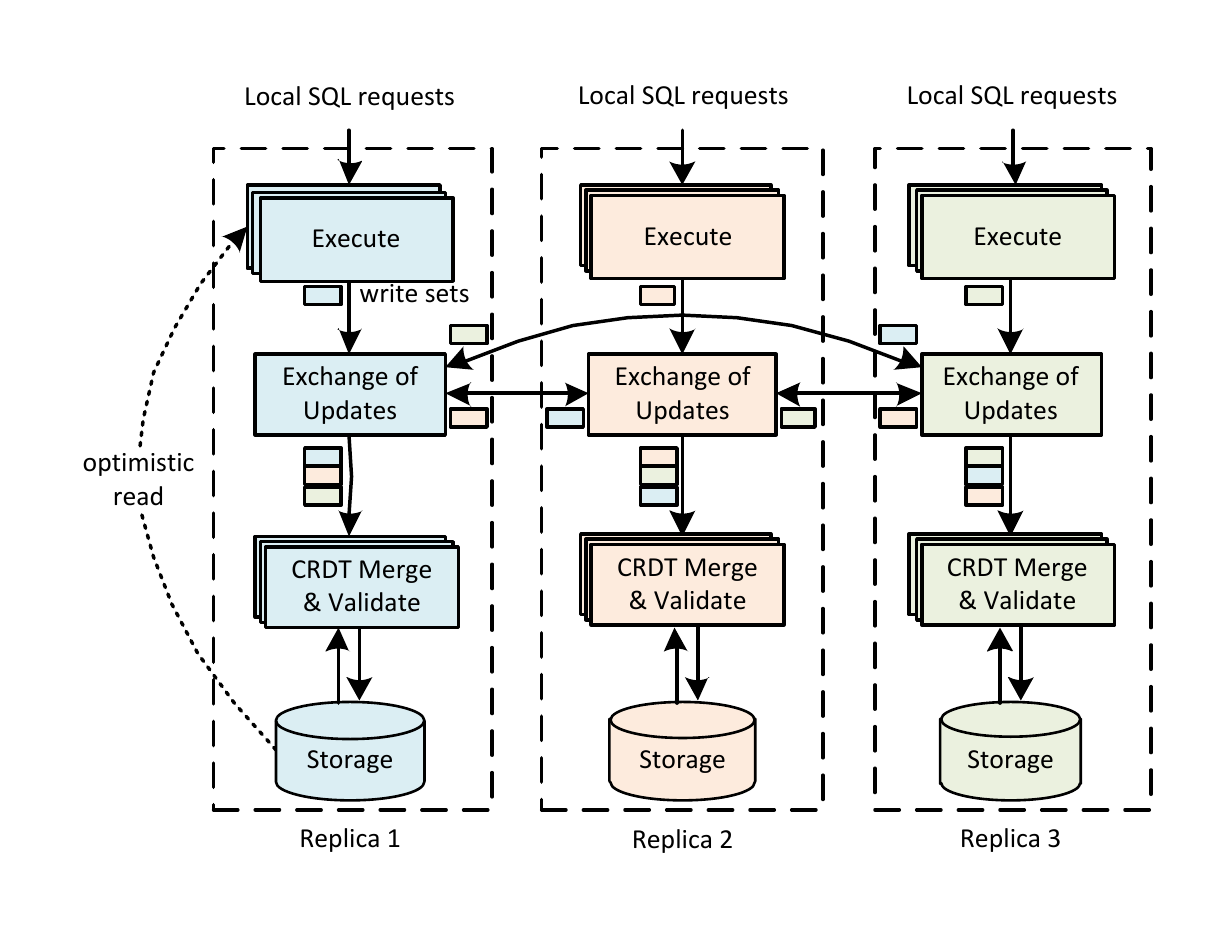}}
  \centering
  \vspace{-0.25in}
  \caption{Overall structure.}
  \label{fig:overall}
  \vspace{-0.2in}
\end{figure}

This section provides system overview of \oursys. The overall architecture is shown in Figure \ref{fig:overall}.
\oursys is developed based on a relational database openGauss \cite{opengauss}. We rely on its SQL engine to parse SQL statements and generate physical execution \red{plans} and use its row-based memory-optimized storage engine openGauss MOT \cite{mot} to store data 
\blue{(other in-memory row-stores with concurrency support are possible alternatives)}. 
The local SQL requests are optimistically executed based on the local current database state. Only the write sets (\aka updates) obtained after local SQL execution are exchanged between nodes, which will be merged with local updates at the receiver side.
The merged updates are then applied to each local replica. 
Some key features are described as follows.

\Paragraph{Multi-Master Architecture with Full SQL Engine}
{\oursys} provides a full SQL engine, supporting standard SQL, application programming interfaces, interoperability, {\etc} {\oursys} is deployed on multiple regional servers, each accepting local users' SQL requests and converting high-level SQL statements to low-level read and write requests through the parser, optimizer, and execution engine.
Each regional server processes its local SQL requests independently based on its local replica.

\Paragraph{Optimistic Read on Local Replica} Read-only transactions are directly served with a snapshot read result for low latency. For transactions that contain both read and write operations, we perform optimistic read on the most recent consistent local replica. If the read set of a transaction becomes stale which is discovered before data replication, the transaction is aborted according to different isolation levels. The optimistic read on local replica helps improve performance under weak isolation levels.

\begin{figure}[h]
\vspace{-0.3in}
  \centerline{\includegraphics[width=4in]{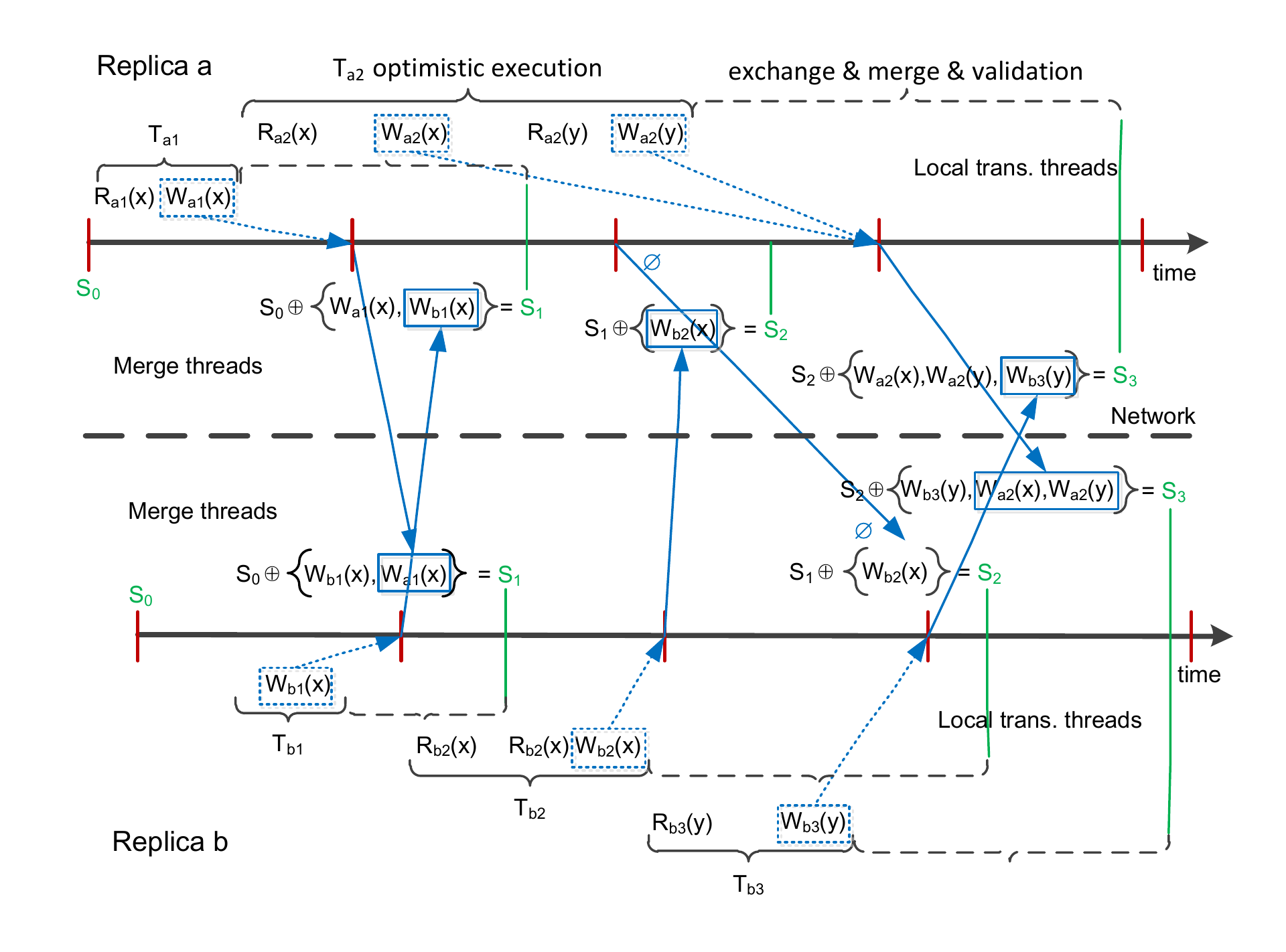}}
  \centering
  \vspace{-0.25in}
  \caption{An illustrative example of update merge with two replicas. Each replica $a$ independently accepts local transactions, \eg $T_{a1}$ and $T_{a2}$. $R_{a1}(x)$ and $W_{a1}(x)$ represent the read and write operations on data item $x$ of replica $a$, respectively. A local transaction with commit epoch number $\cen$ cannot be confirmed until snapshot $\cen$ is generated, so there exists a period for receiving remote updates, merge, and validation. Red bars are epoch boundaries determined by the local clocks, which are not necessarily synchronized across replicas. Green $S_i$ is the globally consistent snapshot. The blue dotted/solid box is the sent/received write set. }
  \vspace{-0.2in}
  \label{fig:rc}
\end{figure}

\Paragraph{Epoch-based Replication of Updates}
All master nodes exchange their local updates periodically (every epoch). The \textit{epoch} is a short period of time (\eg 10 ms), and the epoch number is monotonically increased according to local physical time. Different from deterministic databases that perform replication of the SQL statements, \oursys exchanges the \textit{write sets} between masters, which can be considered as the \textit{delta state} in state-based CRDT merge. 
We provide an illustrative example in Figure \ref{fig:rc} showing how it works under two replicas setting. 
Suppose a consistent snapshot $S_0$ among two replicas in the beginning, replica $a$ and replica $b$ independently execute their local transactions and independently generate their local write sets. For the transactions that finish execution, their write sets (\aka updates) are exchanged between replicas at the end of every epoch. After a node receives all remote updates committed at the first epoch, these updates are merged with local write sets to resolve write-write conflicts based on a \textit{CRDT merge rule}, which guarantees the same merge result given the identical set of updates \textit{regardless of arriving order}. 
On each node, the identical merge outcome is applied to the original snapshot $S_0$ to generate a new snapshot, \ie $S_1=S_0\oplus\{W_{a1}(x), W_{b1}(x)\}$, where $\oplus$ represents the merge operation. Only when the local snapshot for epoch $i$ is generated (after merge and validation), can the local transactions with commit epoch number $i$ be returned by the host master with committed/aborted notification, \eg $T_{a2}$ which is committed at epoch 3 will not return until $S_3$ is generated on replica $a$, at which time replica $a$ has received and merged the remote updates that are committed at epoch 3, $W_{b3}(y)$.

\Paragraph{CRDT Merge with Transaction Processing} 
By only exchanging write sets (\aka updates), the merge of updates and concurrent transaction processing can be unified by our \textit{multi-master OCC} algorithm, which is performed on a per-epoch basis.
From each replica's perspective, a globally consistent snapshot $i$ will be achieved once it has merged the updates of epoch $i$ and all previous epochs (collected from local and all other remote nodes), at which time the local transactions of epoch $i$ can be returned with committed/aborted response. \blue{We rely on delta-CRDT to efficiently resolve cross-region conflicts and achieve a globally consistent state. With a merge operation that has ACI properties, these writes from different replicas can be partially merged to improve efficiency (Associative), can arrive in different orders (Commutative), and even can be retransmitted (Idempotent). The replica state after merging local/remote updates is guaranteed to be \textit{consistent} among all nodes. }

\Paragraph{Asynchronous Execution and Synchronous Validation}
A node is allowed to execute transactions of epoch $i$ even though snapshot $(i-1)$ has not been generated. In other words, the execution is not necessarily synchronized across epochs, but the validation has to be synchronized. The validation of epoch $i$'s transactions is performed based on the globally consistent snapshot $(i-1)$, where the previously executed transactions that have write-write conflicts with others should be aborted (by checking whether a transaction's pre-write is overwritten by others). As a conventional way, data replication and transaction processing are designed separately, losing the opportunity for achieving high concurrency. We support high-throughput transaction processing by coupling the merge of updates and the optimistic transaction processing, which will be discussed in Section \ref{sec:4}.

%% file: trans.tex
\section{Epoch-based Multi-Master OCC}
\label{sec:4}
{
\begin{algorithm}[t]
\SetAlgoNoLine
\small
    \caption{Local Transaction Process (a Thread per Transaction)}
    \label{alg:tx}
    
    \KwIn{a transaction $T$}
    \KwOut{return commit or abort}
    \{$T.\sen, T.\lsn\}\leftarrow$ get current epoch no. and latest snapshot no.\;
    Execute transaction $T$ based on latest snapshot\;
    $T.RS\leftarrow$ $T$'s read set; $T.WS\leftarrow$ $T$'s write set\;
    $\{T.\csn, T.\cen\}\leftarrow$ get current timestamp and epoch no.\;
    \uIf {$Isolation == \RC$}{
         Add $T.\{\sen,\csn,\cen,WS\}$ to send buffer\;
    }
    \uElseIf{$Isolation == \RR$ or $\SI$}{
        \textcolor{gray}{//Read set validation}\\
    	\For{\rm{each} $record$ \rm{in} $T.RS$}{
    	   $row = FindRow(record.key)$\;
    	   \uIf{$row == null$}{
    	       Abort $T$; \textcolor{gray}{ \phantom{aa}//read row is deleted} 
    	   }
    	   \uElseIf{$\RR$ and $record.\csn \neq row.\csn$}{
    	       Abort $T$; \textcolor{gray}{ \phantom{aa}//read row is updated} 
    	   }
    	   \uElseIf{$\SI$ and $row.\cen-1 > T.\lsn$}{
    	       Abort $T$; \textcolor{gray}{ \phantom{aa}//snapshot is updated} 
    	   }
        }
    }
    \uElse{
        \textcolor{gray}{//Not support}\\
    }
    \uIf{$T.WS==\emptyset$}{
        return COMMIT; \textcolor{gray}{ \phantom{aa}//read-only transaction} 
    }
    Add $T.\{\sen,\csn,\cen,WS\}$ to send buffer\;
    \underline{wait till snapshot $T.\cen-1$ is generated;}\\
    \texttt{DeltaCRDTMerge}(T.\{\sen,\csn,\cen,WS\}); \textcolor{gray}{\phantom{aa}//Algorithm2, based on the consistent snapshot $(T.\cen-1$)}\\  
    \uline{wait till all updates of remote/local TXs with $T.\cen$ are applied on rowheaders;}
    \\[3pt] 
    \textcolor{gray}{//Validation:}\\
    \For{\rm{each} $record$ \rm{in} $T.WS$}{
        $row = FindRow(record.key)$\;
        \uIf{$row.\csn\neq T.\csn$} {
            Abort $T$; \textcolor{gray}{ \phantom{aa}//write-write conflict occurred}
        }
    }
    \textcolor{gray}{//Write-back:}\\
    \For{each $record$ in $T.WS$}{
        $row = FindRow(record.key)$\;
        $row.write\_data(record.data)$\;
    }
    return COMMIT\;
\end{algorithm}
}
\subsection{Lifecycle of a Local Transaction} 
A transaction is submitted from the client to the local server, where a \textit{thread} is assigned for the transaction. \textbf{Algorithm \ref{alg:tx}} describes the per-thread transaction processing logic. A transaction is first assigned with a \textit{start epoch number} ($\sen$) and \red{a} \textit{latest snapshot number} ($\lsn$) (Line 1). The $\lsn$ is the latest globally consistent snapshot number maintained by the server at the current time, which is used for read set validation in snapshot isolation. The transaction is then executed and generates read set $RS$ and write set $WS$ (Line 2-3). The SQL constraints are checked during the transaction execution. If a constraint violation occurs, the transaction will abort before generating the write set. In the case of a read-only transaction, it reads data on the most recent snapshot and returns after read validation(Line 19-20).

For the transactions that contain write operations, our multi-master OCC requires each transaction to record a few meta information for CRDT merge. Specifically, a transaction is assigned with a \textit{commit epoch number} ($\cen$) and a \textit{commit sequence number} ($\csn$) (Line 4). The $\cen$ is the current physical epoch number used to determine the batch of transactions that attempt to commit together, including both the local and remote transactions with the same $\cen$. The timestamp along with its local server ID is used to generate a globally unique $\csn$, which is used to determine the execution order within the same epoch. In addition, the transaction's write set $WS$ along with its $\{\sen,\csn,\cen\}$ is sent to remote peers. \oursys supports several ANSI isolation levels, \eg \textit{Read Committed} ($\RC$), \textit{Repeatable Read} ($\RR$), and \textit{Snapshot Isolation} ($\SI$). They are different in processing logic (Line 5-18).

For a transaction with $T.\cen$, a synchronization point is placed to wait for snapshot ($T.\cen-1$) to be generated (Line 22). The snapshot ($T.\cen-1$) is generated by merging all local and remote transactions of epoch ($T.\cen-1$). This warrants that operations are applied on the most recent consistent snapshot, otherwise it could impact correctness and consistency (Section \ref{sec:4:consistency}). 

The \texttt{DeltaCRDTMerge} function is then launched (Line 23), which defines the rule for merging a new update (\aka delta state) into the current database state (see Section \ref{sec:4:trans:update} and Algorithm \ref{alg:precommit}). A \textit{row header} stores each row's meta information $\{\sen, \lsn, \csn, \cen\}$ indicating the row's update history by an arbitrary transaction thread, which is used for OCC's \textit{pre-write}. For example, suppose a row header is firstly updated by a transaction $T$, \ie update row header by $row.\csn=T.\csn$, is then overwritten by other threads leading to $row.\csn\neq T.\csn$, $T$ will be aborted during the validation phase (Line 26-29). This implies that there exists a write-write conflict on the same row. According to the merge rule, only one write wins and is committed, and the others are aborted. 

The validation phase of transaction $T$ cannot start until all local/remote transactions of epoch $T.\cen$ are collected and applied on the row headers (Line 24). This is essential to the correctness by ensuring none of the updates are missing. If the transaction does not meet a write-write conflict or wins in conflict merge, it is allowed to commit. This transaction's write data are used to update the involved rows (Line 31-33). After all transactions with the same commit epoch number $\cen$ are validated and applied on the table, a new snapshot for epoch $\cen$ is generated.

\subsection{Merge of Updates} 
\label{sec:4:trans:update}
\begin{algorithm}[t]
\SetAlgoNoLine
    \caption{DeltaCRDTMerge}
    \label{alg:precommit}
    \small
    \KwIn{a transaction's $T.\{\sen,\csn,\cen,WS\}$, current row headers}
    \KwOut{updated row headers}
    \For{\rm{each} $record$ \rm{in} $T.WS$}{
        $row = FindRow(record.key)$\;
        \uIf{$row == null$}
        { 
            Abort $T$; \textcolor{gray}{ \phantom{aa}//row is deleted by other threads} \\
        }
        \uIf{$row.\cen < T.\cen$} 
        {
    	$row.\{\sen,\csn,\cen\} = T.\{\sen,\csn,\cen\}$; \textcolor{gray}{\phantom{aa}//row is not pre-written in current epoch} \\
        }
        \uElseIf{$row.\cen == T.\cen$}
        {
            \uIf{$row.\sen==T.\sen$}
            {
                \uIf{$row.\csn > T.\csn$} 
                { 
                    $row.\{\sen,\csn,\cen\} = T.\{\sen,\csn,\cen\}$; \textcolor{gray}{\phantom{aa}//first write wins}\\
                }
                \uElse{
                    Abort $T$; 
                } 
            }
            \uElseIf{$row.\sen < T.\sen$}
            { 
                $row.\{\sen,\csn,\cen\} = T.\{\sen,\csn,\cen\}$; \textcolor{gray}{\phantom{aa}//shorter transaction wins}\\
            }
            \uElse{
                Abort $T$; 
            }
        }
        \uElse{
            \textcolor{gray}{ //$row.\cen > T.\cen$ will never happen}
        }
    }
\end{algorithm}

\begin{algorithm}[t]
\SetAlgoNoLine
    \caption{Epoch-based Multi-Master OCC}
    \label{alg:remote}
    \small
    \KwIn{the continuously received batches of remote updates $TS= \{T.\{\sen,\csn,\cen,WS\}\}$, the set of local transaction processing threads $LT[i]$ for epoch $i$, and the number of replicas $n$}
    \KwOut{continuously updated table}
    \BlankLine
    \underline{\textbf{Receive thread:}}\\
    \While{true}{
         $TS\leftarrow$ receive a batch of updates with epoch $TS.\cen$;\\
         $BUF[TS.cen]$.add($TS$);\textcolor{gray}{ \phantom{aa}//add TS to buffer $BUF$ indexed by \cen}\\
    }
    \BlankLine
    \underline{\textbf{Merge thread:}}\\
    \While{true}{
        $\lsn\leftarrow$ get latest consistent snapshot no.;\\
        $TS$ = $BUF[\lsn+1]$.get();\textcolor{gray}{ \phantom{aa}//blocking get}\\
         \ForEach{$T$ in $TS$}{
             \texttt{DeltaCRDTMerge}($T.\{\sen,\csn,\cen,WS\}$);\\
             \uIf{$T$ is not aborted}{
                 $Q[\lsn+1]$.push($T$); \textcolor{gray}{ \phantom{aa}//add $T$ to commit queue $Q$}\\
             }
         }
         $N[\lsn+1]\leftarrow N[\lsn+1] + 1$;\textcolor{gray}{ \phantom{aa}//counts the executed remote $TS$}\\
         \uIf{$N[\lsn+1]==n-1$}{
             \textcolor{gray}{ //all remote updates have been processed}\\
             NotifyAll($\{Thread[T]\mid T.{\cen}=\lsn+1\}$);\textcolor{gray}{ \phantom{aa}//notify all threads blocked at Line 24 (Alg. 1)}\\
             \ForEach{$T$ in $Q[\lsn+1]$}{
                 Execute Line 26-33 of Alg. \ref{alg:tx} for $T$;\textcolor{gray}{ \phantom{aa}//validate and write}\\
             }
             \underline{wait till all local threads($Thread.T.cen == \lsn + 1$) are finished;}\\
             NotifyAll($\{Thread[T] \mid T.{\cen}=\lsn+2\}$);\textcolor{gray}{ \phantom{aa}//notify all threads blocked at Line 22 (Alg. 1)}\\
             $\lsn\leftarrow \lsn+1$;\textcolor{gray}{ \phantom{aa}//a new snapshot is generated}\\
         }
    }
    \BlankLine
    \underline{\textbf{Local transaction processing primary thread:}}\\
    \While{true}{
        Fork a new thread $Thread[T]$ for a local transaction $T$;\\
        $Thread[T]$.start();\textcolor{gray}{ \phantom{aa}//Algorithm 1}\\
        
    }

\end{algorithm}

\setlength{\floatsep}{10.0pt plus 2.0pt minus 2.0pt}
\setlength{\textfloatsep}{10pt plus 1.0pt minus 2.0pt}
\subsubsection{Epoch-Aware Delta CRDT Merge} 

CRDTs \cite{shapiro2011comprehensive,Shapiro:2011:CRD:2050613.2050642} are distributed datatypes that allow different replicas of a distributed CRDT instance to diverge and
\red{ensure} that, eventually, all replicas converge to the same state. State-based CRDTs achieve this by propagating updates of the local state by disseminating the entire state across replicas. The received states are then merged to
remote states, leading to convergence (\ie consistent states on all replicas). In delta CRDT \cite{almeida2018delta}, only the updates are disseminated instead of the entire state. A received delta-state (\ie update) is incorporated with the local full state (\ie database) via a merge function that deterministically reconciles both states. The delta-states can be shipped using an unreliable dissemination layer that may drop, reorder, or duplicate messages, \ie delta-states can always be out of order, re-transmitted, and re-joined. Furthermore, it is desired to handle concurrent updates for high performance. Therefore, the key is the design of the merge function which must be associative, commutative, idempotent, and parallel-friendly, and most importantly be aware of epoch information.

Our merge function, \texttt{DeltaCRDTMerge}, is depicted in \textbf{Algorithm \ref{alg:precommit}}. 
It defines the conflict merge rule to determine the successful updates and may be concurrently invoked by multiple local transactions and remote transactions. It relies on a data structure, \textit{row header}, to support epoch-based delta-state merge, which stores each updated row's $\{\sen, \lsn, \csn, \cen\}$. For ease of illustration, we focus on the \texttt{update} transactions. 1) If a write row is null, which means that it was deleted by other threads in past few epochs, we abort this transaction (Line 3-4). 2) If the write row is not pre-written in the current epoch ($T.\cen$) yet, \ie $row.\cen < T.\cen$, we update its row header for a candidate commit (Line 5-6). Otherwise,
this row has been updated in the current epoch by other threads, \ie $row.\cen == T.\cen$. Note that, since the \texttt{DeltaCRDTMerge} will never be invoked to merge updates of epoch $(i+1)$ before completing the merge of all updates of epoch $i$ (the synchronous point at Line 22 in Algorithm \ref{alg:tx}), $row.\cen > T.\cen$ will never happen. 3) We let shorter transactions win by comparing their $T.\sen$, \ie $row.\sen < T.\sen$. A transaction with larger $T.\sen$ means that it is closer to the current epoch $T.\cen$ (because $T.\cen-T.\sen$ is smaller) and is likely to commit (Line 13-14). 4) Suppose the same $\sen$, we use $\csn$ to determine the order and follow the first-write-win rule (Line 8-12). Note that, for an insert request, the \texttt{FindRow} function cannot locate the row. In addition, multiple concurrent insert transactions may insert into the same row. We use a \textit{temporary table} to store the inserted rows to deal with the insertion conflicts within the same epoch. Instead of invoking \texttt{FindRow} \red{which} relies on the table index, we use a temporary table for the inserted rows.

\blue{The \texttt{DeltaCRDTMerge} operation defined in \textbf{Algorithm \ref{alg:precommit}} is similar to the first-write-win conflict resolution in OCC, but they have different design goals. Our merge function that has ACI property is designed for merging the local/remote updates on separate nodes to achieve a globally consistent state (\ie replicating states), even though these updates arrive at each node in different orders (Commutative property), are merged partially (Associative property), are retransmitted (Idempotent property). But the first-write-win rule in OCC is for deciding the committed and aborted transactions in concurrent transaction processing on a single replica.}

\subsubsection{Handling Remote Updates} 
The merge of remote updates and the local transaction processing are running concurrently and \red{interacting} with each other. On each master node, multiple receive threads and merge threads are continuously running as shown in \textbf{Algorithm \ref{alg:remote}}. The receive thread keeps receiving updates $TS$ from remote peers and temporarily stores them in the receive buffer with their commit epoch number $T.\cen$ information (Line 3-4). Suppose $\lsn$ is the number of the most recent globally consistent snapshot (Line 7). The merge thread merges updates of epoch ($\lsn+1$) based on the most up-to-date table state (\ie snapshot $\lsn$) and will produce consistent snapshot \textit{one by one}. Specifically, if there exist remote updates of epoch ($\lsn+1$) in the buffer, it processes them by invoking the \texttt{DeltaCRDTMerge} function (Line 8-10). The transactions that are not aborted are pushed into a commit queue $Q[\lsn+1]$ (Line 11-12). If the updates of epoch ($\lsn+1$) from all remote peers have been applied on row headers, it notifies all the local transaction threads for epoch ($\lsn+1$) and triggers the validation and write-back for each update in the commit queue (Line 16-18). Once all the local threads of epoch ($\lsn+1$) are finished, it means that a new consistent snapshot is achieved. The merge thread immediately notifies all the local transaction threads of the next epoch that are waiting for this up-to-date snapshot (Line 19-21). We also have a primary thread for assigning a new thread for each new local transaction (Line 23-25).

\subsubsection{Case Study} 
\Paragraph{1) Long running Transactions} Only when a transaction commits at epoch $i$ can we send its write sets at the end of epoch $i$. For example in Figure \ref{fig:rc}, $T_{a2}$ is a long running transaction that crosses three epochs and ends at epoch 3. 
If the transaction is not aborted in the read validation phase, its write sets $\{W_{a2}(x),W_{a2}(y)\}$ are sent out together at the end of epoch 3. 
\blue{Note that, the transactions that start at the same epoch are not necessarily completed synchronously before the next epoch can start. They may finish execution and commit at different epochs, so the short transactions do not wait for the long transactions but can commit at earlier epochs. A long transaction (spanning multiple epochs) is more likely to be aborted due to inconsistent read in \RR or stale read in \SI, where the read data might be modified by early committed short transactions.}
Nevertheless, if there are no updates for an epoch, \eg epoch 2 on replica $a$, an empty message is sent out to prevent endless wait at remote peers.  
\textbf{2) Network Delay.} For example in Figure \ref{fig:rc}, due to network latency, the two replicas probably may not reach a consistent snapshot exactly at the same time point, but they will at a certain time. This will NOT block the system running and will NOT impact sequential consistency (see Section \ref{sec:4:consistency}). For example, snapshot $S_2$ is generated at epoch 3 on replica $a$ and at epoch 4 on replica $b$. The write set $W_{b3}(y)$ of $T_{b3}$ is routinely sent out to replica $a$ at the end of epoch 3. 
However, the merge of updates of epoch $i$ cannot start until snapshot $(i-1)$ is generated (Line 22 in Algorithm \ref{alg:tx}). For example, local update $W_{b3}(y)$ of epoch 3 cannot be merged with snapshot $S_1$, but can only be merged with snapshot $S_2$ and the updates of epoch 3 ($W_{a2}(x)$ and $W_{a2}(y)$) to generate $S_3$.

\subsection{Isolation}
\label{sec:4:isolation}

Weak isolation models cannot guarantee serializability, but their benefits to concurrency are frequently considered by application developers to outweigh the costs of possible consistency anomalies that might arise from their use. \oursys supports multiple weak ANSI isolation levels as shown in \textbf{Algorithm \ref{alg:tx}}, allowing users to choose the optimal one for their specific application.

\Paragraph{Read Committed (\RC)} For a transaction $T$ of epoch $T.\cen$, only the committed data as a snapshot can be accessed in our system, so the support of \RC isolation is straightforward. None of the local transactions is aborted and all of them 
are used to generate write sets. Its write set $T.WS$ along with its metadata are immediately added to the send buffer (Line 21). Note that it cannot guarantee to read the most up-to-date committed data since the most recent snapshot $T.\cen-1$ might not be generated yet. But it can guarantee that the latter read item is a more up-to-date one.

\Paragraph{Repeatable Read (\RR) and Snapshot Isolation (\SI)} As a single-version in-memory database, we realize the \RR and \SI mainly through \red{an} \textit{optimistic} approach, \ie read set validation. 1) If the previously read record is deleted, the transaction is aborted under these two isolation levels (Line 11-12). 2) During a transaction’s execution, each time it reads a record, the row's \csn is updated as the record’s \csn. If a row is updated by other threads after its first read (\ie $row.\csn\neq record.\csn$), the transaction is aborted under \RR isolation (Line 13-14). 3) If a transaction's read row is updated in a new snapshot (snapshot $row.\cen-1$ due to the synchronous point at Line 22) since it starts execution (snapshot $T.\lsn$), the initial read snapshot is updated (\ie $row.\cen-1 > T.\lsn$) which violates \SI. The transaction is aborted under \SI (Line 15-16). Only the transactions that pass read set validation will be sent out (Line 21). 
For example in Figure \ref{fig:rc}, $T_{b2}$ is aborted under \RR because its first read $R_{b2}(x)$ is the $x$ in $S_0$ while its second read $R_{b2}(x)$ is the updated $x$ in $S_1$. With \SI, a transaction is aborted if its read snapshot is stale. For example, $T_{a2}$ will be aborted under \SI because its first read $R_{a2}(x)$ is based on $S_0$ while its second read $R_{a2}(y)$ is based on $S_1$.

\Paragraph{Serializable Snapshot Isolation (SSI)} We discuss the possibility to support SSI \cite{cahill2009serializable}. We need to exchange each read record's key for read-write dependency detection. Then we can prevent snapshot isolation anomalies by aborting a transaction when a pair of consecutive conflict edges are found. Currently, \oursys do not support SSI due to its high cost for transferring the read keys.

\subsection{Consistency of Replicas}
\label{sec:4:consistency}

We show how \oursys guarantees the consistency of replicas.

\vspace{-0.05in}
\begin{lemma}
\label{lemma:read}
Following epoch-based multi-master OCC, the read operations will not affect the consistency of replicas.
\end{lemma}
\vspace{-0.05in}
\begin{proof}
In our system, each master node performs a local read on snapshots and then generates the write set. That is, the write set is generated by a single source worker and then sent out for write-write conflict merge. No matter which snapshot (old or new) a transaction reads, the write data by the transaction are \textit{deterministic} before they are sent out. The consistency of read operations will only affect isolation property. The transactions that read different versions of data violate isolation constraints ({\eg} \RR and \SI) and will be aborted without producing updates. Only the updates can change the state of replicas. Therefore, the read operations will not affect the consistency of replicas.
\end{proof}

\vspace{-0.05in}
\begin{lemma}
\label{lemma:write}
Suppose an initial database state $S_i$ and a set of updates $TS=\{\mathcal{U}(T_1), \mathcal{U}(T_2),\ldots,\mathcal{U}(T_n)\}$ with the same \cen, where $\mathcal{U}(T_i)=T_i.\{\sen,\csn,\cen,WS\}$. No matter what order these updates input in, Algorithm \ref{alg:precommit} will produce the deterministic state $S_{i+1}$.
\end{lemma}
\vspace{-0.05in}
\begin{proof}
The merge rule (Algorithm \ref{alg:precommit}) is defined by comparing tuples $\{\sen,\csn,\cen\}$. We define a strict total order `$\prec$' on the set of updates $TS$ as follows. For any two updates $\mathcal{U}(T_i)$ and $\mathcal{U}(T_j)$ with the same \cen (\ie $T_i.\cen=T_j.\cen$), we have $\mathcal{U}(T_i)\prec\mathcal{U}(T_j)$ if one of the following conditions is met. 
\begin{itemize}
    \item $T_i.\sen > T_j.\sen$;
    \item $T_i.\sen = T_j.\sen$ and $T_i.\csn<T_j.\csn$.
\end{itemize}
Because $T.\csn$ is generated based on $T$'s source worker id and its commit timestamp assigned by the source worker, which is \textit{globally unique}, there must be a strict order between any two updates $T_i$ and $T_j$. For conflict updates on a row $x$, Algorithm \ref{alg:precommit} allows for the updates of transaction $T$ with the ``smallest'' (according to `$\prec$') to commit. That is, the collection of successful updates always equals to $\big\{\{x\in\mathcal{WS}:T_i\in\mathcal{C}(x)\}\mid \mathcal{U}(T_i)\prec\mathcal{U}(T_j),\forall T_j\in\mathcal{C}(x)\big\}$, where $x$ is a table row, $\mathcal{WS}=\{T_1.WS\cup T_2.WS\cup\ldots\cup T_n.WS\}$ is the union of the write sets of all transactions in an epoch, and $\mathcal{C}(x)$ indicates a set of transactions that write row $x$. Therefore, with an initial state $S_i$, no matter what order these updates input in (even for duplicated updates), the \textit{deterministic} order implicitly defined on the set of updates with unique tuples $\{\sen,\csn,\cen\}$ will guarantee the deterministic output $S_{i+1}$.
\end{proof}

\vspace{-0.05in}
\begin{theorem}
The epoch-based multi-master OCC (Algorithm \ref{alg:tx},\ref{alg:precommit},\ref{alg:remote}) enforces consistency of replicas at the granularity of epochs.
\end{theorem}
\vspace{-0.05in}
\begin{proof}
Lemma \ref{lemma:read} states that read operations will not affect the consistency of replicas, so we only focus on studying the effects of write operations. Lemma \ref{lemma:write} states that given \red{a} batch of updates and an initial state, the output state is \textit{deterministic} regardless of the non-deterministic input order of these updates. To achieve consistency of replicas, it is crucial to guarantee 1) the identical initial state and 2) the identical batch of updates on all replicas. First, the synchronization points at Line 24 in Algorithm \ref{alg:tx} and at Line 19 in Algorithm \ref{alg:remote} ensure that each replica runs an \textit{identical} batch of updates, \ie having merged all local/remote updates of an epoch. Second, the synchronization point at Line 22 in Algorithm \ref{alg:tx} and the latest snapshot number verification at Line 7 in Algorithm \ref{alg:remote} ensure that the updates of epoch $(i+1)$ are applied on snapshot $i$, which guarantees the \textit{identical} initial snapshot. Therefore, the consistency of replicas is guaranteed at the granularity of epochs.
\end{proof}
\vspace{-0.05in}
The read operations on the local replica are not exchanged among replicas. The order of write operations in an epoch is deterministic and consistent on each individual replica (Lemma \ref{lemma:write}). The epoch snapshots on all replicas are generated one by one. Thus, the consistency enforced by our algorithm is a kind of \textit{sequential consistency}.

%% file: impl.tex
\section{Implementation}
\label{sec:4:impl}

\begin{figure}
\vspace{-0.3in}
  \centerline{
  \includegraphics[width=2.8in]{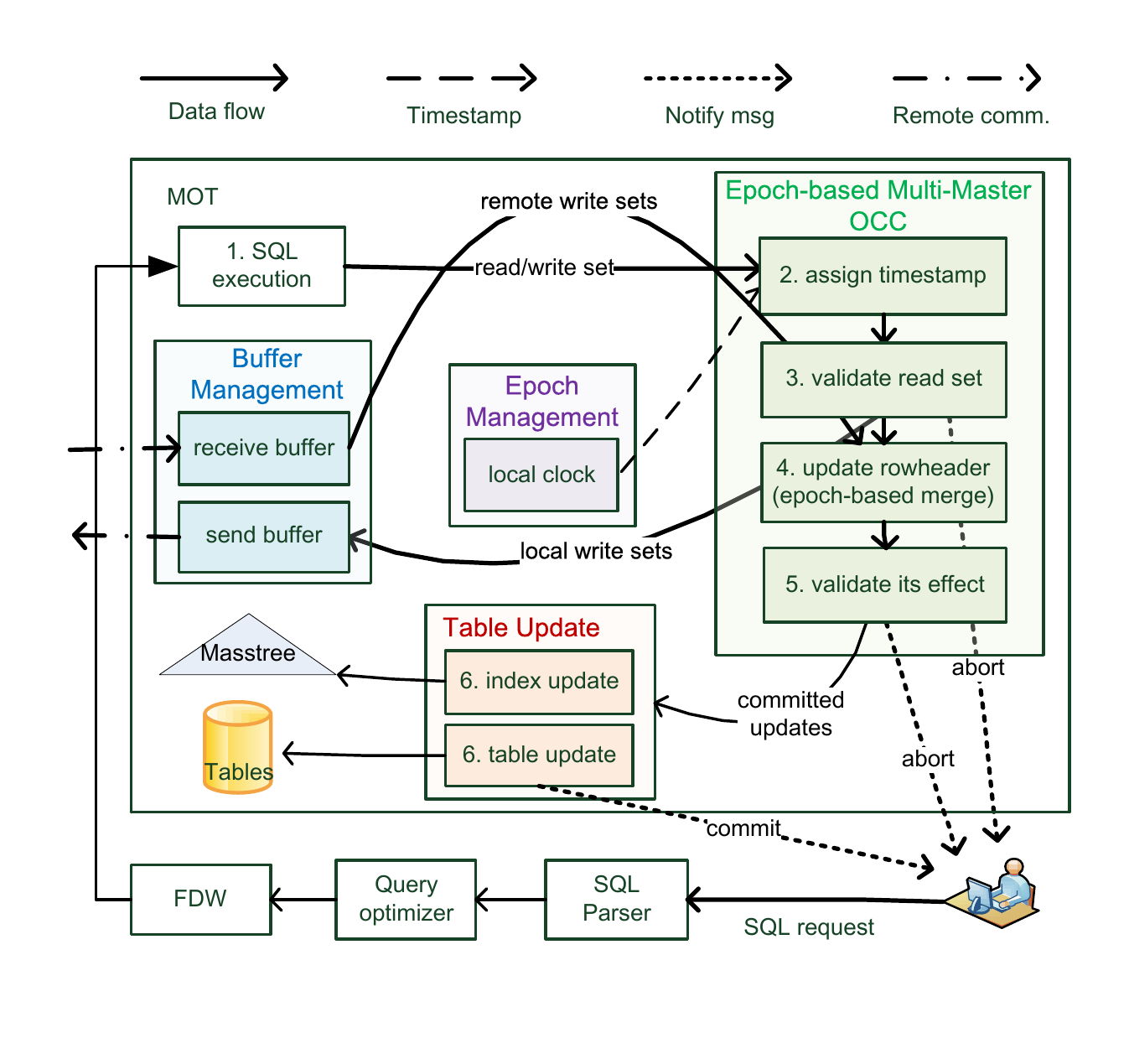}}
  \centering
  \vspace{-0.35in}
  \caption{Data flow of GeoGauss.}
  \label{fig:dataflow}
  \vspace{-0.1in}
\end{figure}

We implement \oursys based on openGauss 2.0 MOT storage engine \cite{opengauss}. The data flow of \oursys is depicted in Figure \ref{fig:dataflow}.

\subsection{Parallelism and Communication}
\label{sec:impl:parallel}

\Paragraph{Thread Blocking and Notification} A large number of transaction threads might be blocked waiting for a notification signal (Line 22 in Algorithm \ref{alg:tx}). It is possible that a thread entering into the blocking phase and a signal notifying the thread concurrently occur. If the thread misses the notification signal, it will be suspended permanently. To avoid the anomalies, we let each thread sleep and actively check the condition periodically, which works well with a small number of threads. On the other hand, we introduce a mutex lock to prevent concurrency anomalies. But these threads that share the mutex lock have to be invoked one by one serially. To maximize parallelism, we create multiple mutex locks and make each one only shared by a disjoint subset of threads. This can effectively alleviate the burst of thread notifications.

\Paragraph{Message Queue and Pipelining} We rely on Protocol Buffers \cite{protobuf} and Gzip \cite{gzip} to compress the buffered write sets. Instead of the heavyweight gRPC \cite{gRPC} used by openGauss, we prefer the lightweight ZeroMQ \cite{zeromq} (with publish-subscribe mode) to realize efficient data transmission between nodes. Furthermore, as depicted in Figure \ref{fig:rc}, the write sets in a send buffer are packaged and sent out together at the end of each epoch. This could seize computation resources and incur network bursts. To overlap the communication and computation, we introduce a pipelining technique that sends write sets in mini-batches in a streamlined manner. Note that, an EOF message is sent indicating the end of the epoch (according to physical time) in streamlined communication, so the receiver can ensure the completeness of an epoch of transactions. Furthermore, we improve the pipelining by zero-copy send and receive. Usually, when a user space process has to execute system operations like reading or writing data from/to a network device, it has to perform one or more system calls that are then executed in kernel space by the operating system. We copy the write sets data through zero-copy and directly send them out through Protocol Buffers. This can alleviate the serialization and deserialization costs.

\vspace{-0.05in}
\subsection{Fault Tolerance}
\label{sec:imple:fault}

\blue{Under a multi-master architecture, once a master node fails to provide service, its local transaction requests can be routed to another master node. We employ Raft protocol \cite{braft} to make a consensus on the status of live nodes, which can prevent permanent blocking (the whole system may be blocked waiting for the write sets from a failed node). This is light weight since it is invoked only when the status of alive nodes is changed. Under such an epoch-consistent replicated system, multiple master nodes work as strongly consistent replicas, so there is little risk that all master nodes fail simultaneously. We can also place more replicas for backup to increase robustness as existing commercial databases do. However, there is a unique problem that should be tackled under such architecture. That is, we should prevent the loss of updates problem, \ie some updates successfully commit on the local node but fails on remote nodes. To avoid such faults, \oursys provides three fault tolerance options, from light to heavy-weighted.}

\Paragraph{\blue{Local WriteSet Backup Server}} 
\blue{We place one (or more) writeset backup servers associated with each replica in each region, caching the generated local write sets for backup. Each time a server node sends local write sets to remote nodes, it also sends a copy to the local writeset backup server, which will send back an ack message as a response. Only when all remote write sets with commit epoch number $\cen$ have been collected and the local write sets with $\cen$ have been backed up, can we return to users the committed/aborted states of the transactions with $\cen$. Once a node failure is detected and removed, the other remote nodes will ask the corresponding writeset backup server to check whether there exists a loss of updates by comparing the monotonic $\cen$. If so, the remote nodes need to pull the lost write sets and merge them to advance to the consistent snapshot $\cen$ before proceeding. Since the local backup occurs simultaneously with the sending of local updates and the local network is much faster than the cross-region network, the cost of local backup is hidden, which will not impact the performance.}

\Paragraph{\blue{Remote WriteSet Backup Server}}
\blue{In case a region fails, the local backup server will fail together, and the above scheme may not work. Similarly, we can place one (or more) remote backup servers in other regions for caching the write sets. In this case, it requires one cross-region round-trip-time (RTT) for sending backup updates and receiving the ack message. This is more expensive than the local backup method, which requires only 0.5 cross-region RTT for receiving the remote updates. This solution can ensure fault tolerance when one or more regions fail.}

\Paragraph{\blue{Raft Replication of WriteSets}}
\blue{To maximize the fault tolerance, the write sets can be replicated through Raft consensus protocol on a per-epoch basis. This is similar to the deterministic databases which rely on Raft to replicate SQL inputs. Each node (as a Raft leader) sends the generated write sets to other receiver master nodes (as Raft followers), who will send back ack messages. If the leader collects more than half of the ack messages, it sends a commit request to the receivers indicating that the write sets can be applied to the existing database state. Therefore, the receiver requires $\thicksim$1.5 RTTs to receive the write sets from the sender node.
}

%% file: discuss.tex
\vspace{-0.05in}
\section{Discussion}
\label{sec:discuss}

\Paragraph{Transaction-based vs. Batch-based vs. Epoch-based}
Traditional wisdom prefers to handle transactions on a per-transaction basis \cite{corbett2013spanner,taft2020cockroachdb}. This is suboptimal under \red{a} high-contention workload and geo-distributed environment due to the expensive coordination cost for each individual transaction. Our multi-master OCC algorithm can also be slightly changed to support transaction-based conflict merge. However, this brings more complexity for ensuring the atomicity and consistency of replicas. An alternative is batch-based processing (\eg deterministic databases \cite{thomson2012calvin,lu2020aria}) that limits the number of updates for each batch instead of \red{a} fixed time period. \blue{The batch-based approach could have undesirable performance due to imbalanced workloads among transactions, especially for long-running transactions, due to the barriers across batches. Our epoch-based multi-master OCC divides transactions into multiple individual subsets according to the local time and their commit time ({\ie} \cen). The updates of long running transactions will be merged in later epochs without blocking the merging process.}
In addition, the epoch-based and batch-based approaches are less expensive for fault tolerance since the consensus and durability are established on a set of transactions instead of an individual transaction. 

\Paragraph{Epoch Length}
Intuitively, the epoch length should be set longer than the round trip time (RTT). But it is not necessary due to the deterministic execution with the coordination-free property. Unlike the coordination-based approach (\eg 2PC) that relies on the remote server's confirmation on the validity of remote data (\eg through locking), our multi-master OCC only verifies the completeness of an epoch of updates (\ie have collected updates from all peers from the receiver's point of view) and does not need to coordinate with remote peers (with one or more RTTs) before proceeding. In our cross-region experiments, it is common that the latest snapshot that is used for generating read/write sets lags behind the current physical time by \textbf{3-5 epochs} (corresponding to the single-trip delay 30-50ms). Despite the epoch length can be set regardless of network RTT, it should be limited by our serving model. Our system reuses the serving model of openGauss, in which a thread is allocated to serve a submitted transaction and will not release its resources until the transaction is committed or aborted. Before that, these serving threads might be blocked. Setting a short epoch length can increase the frequency of update exchanges and as a result shortens the confirmation time, which not only reduces the blocking time to improve throughput but also decreases the latency. However, too frequent communications and update merges will increase the scheduling cost that outweighs the benefits (see Section \ref{sec:5:epoch}). 

\Paragraph{Advantages over Deterministic Databases}
Existing \textit{deterministic databases} \cite{thomson2010case, thomson2012calvin, ren2014evaluation, abadi2018overview, 10.14778/3342263.3342647, 10.14778/3446095.3446098, fauna, lu2020aria, thomson2015calvinfs, qadah2020q} replicate batches of SQL transaction requests to multiple master nodes. After collecting all transactions of the same batch, each replica is required to execute these transactions according to a predefined serial order. This requirement is stricter than that required for an execution to be serializable (which only requires that transactions execute according to \textit{some} serial order) since the operating system schedules threads in a fundamentally nondeterministic way \cite{abadi2018overview}. \blue{This might require a locking mechanism to achieve concurrency while ensuring deterministic serial order, but this results in high scheduling overhead. A number of approaches are proposed to reduce the scheduling overhead, \eg Aria \cite{lu2020aria}, QueCC \cite{qadah2018quecc}, PWV \cite{faleiro2017high}, and LADS \cite{yao2016exploiting}.}
However, there exist several disadvantages of deterministic databases, including 1) lack of support for interactive SQL because it is required to disseminate the SQL statements before executing them, which limits their application; 2) the additional scheduling overhead for determinism; 3) undesirable performance due to imbalanced workloads among transactions, especially for long running transactions (because a previous batch of transactions must finish executing before a new batch can begin). 
Compared with deterministic databases which exchange SQL statements, \oursys which disseminates replica-generated write sets has distinct advantages. 1) We support a full SQL engine (\eg interoperability) by disseminating the write sets when a transaction commits. 2) With our CRDT merge rule for write sets (Algorithm \ref{alg:precommit}), we do not need additional scheduling overhead for ensuring the determinism for concurrent processing. 3) In \oursys, transactions are optimistically executed and arranged into batches \textit{according to their commit time} for synchronous validation. It is allowed to process new transactions during the execution of long transactions, so the impact of long transactions is greatly alleviated.

%% file: expr.tex
\section{Evaluation}
\label{sec:5}

This section evaluates \oursys through cross-region experiments. 

\Paragraph{Cluster Setup} Our cross-region cluster contains 3 geo-distributed nodes (corresponding to 3 replicas), including a node in Zhangjiakou city (North China), a node in Chengdu city (Southwest China), and a node in Shenzhen city (South China). Each node (Aliyun ecs.r6e.8clarge instance) is equipped with 32 vCPUs, 256GB DRAM, running Centos 7.6 OS. The network bandwidth between cross-region nodes is about 100 Mbps/s. A separate node (ecs.c6.8xlarge) is set in each region to simulate the local client for sending SQLs.

\Paragraph{Competitors and Configurations} We choose CockroachDB (\crdb) \cite{taft2020cockroachdb}, 
\blue{\calvin \cite{thomson2012calvin}, \aria \cite{wu2019anna}, \calvinfs \cite{thomson2015calvinfs}, \qstore \cite{qadah2020q}, \slog \cite{10.14778/3342263.3342647}}, and a coordination-free KV database \anna \cite{wu2019anna} for comparison. To investigate the performance improvement by \textit{optimistic asynchronous execution and synchronous validation}, we also implement two variants \oursyss and \oursysa based on \oursys. 
\begin{itemize}[leftmargin=*]
    \item \crdb: For fairness, \crdb is configured with \textit{in-memory store} and configured with \textit{stale reads} from outside the read row's home region.
    As suggested by \crdb documentation, we place 2 additional nodes in each region for maximizing its performance. 
    \crdb supports strong consistency and serializable isolation.
    \item \calvin, \aria, \calvinfs, and \qstore: \calvin and \aria are two typical deterministic databases with multi-master replication. They replicate SQLs instead of write sets, and they do not provide a full SQL engine, so they do not support interactive queries. They provide strong consistency and serializable isolation. \calvin leverages ordered locks to achieve concurrency, and \aria relies on dependency analysis and transaction reordering. For these two systems, we use the implementations from \cite{ariacode} and follow their default configurations. \blue{\calvinfs is a distributed file system that extends \calvin to achieve metadata management. We use the implementation from \cite{CalvinFScode} and follow its default configurations. \qstore, which is implemented based on \calvin, uses a queue-oriented transaction processing method instead of ordered locks to reduce the scheduling overhead.}
    \item \blue{\slog: \slog adopts the sharded master-follower architecture. The input SQLs are contained in logs and are replicated to followers. For cross-shard transactions, they are sent to a single node for deciding a global order, which is used for deterministic execution on each shard replica. The shard replica replays the received logs through deterministic execution to achieve serializability.}
    \item \anna: \anna achieves wait-free execution via the merge of lattice-based composite data structures (similar to CRDT). It only supports causal consistency and eventual consistency (by default), so a submitted SQL request is not returned with a committed or aborted response. It also supports a variety of weak isolation levels, where \RC isolation is configured by default. 
    \item \oursys: Our system is configured with 10ms epoch length by default. It supports strong consistency and multiple weak isolation levels (\RC by default). \crdb and \oursys use a standard benchmark interface, in which each connection only sends a single transaction once at a time. In these two systems, each node is configured with 256 connections.
    \item \oursyss: It adopts \textit{synchronous execution and synchronous validation}, \ie it does not start the execution of epoch $i$'s transactions until the snapshot $(i-1)$ is generated. 
    \item \oursysa: It removes epoch concept and adopts \textit{asynchronous execution and asynchronous validation}. Similar to \anna, it provides eventual consistency and does not guarantee strong consistency at the granularity of epochs. 
\end{itemize}
\vspace{-0.05in}
\Paragraph{Workloads} We use two popular benchmarks, YCSB \cite{10.1145/1807128.1807152} and TPC-C \cite{tpcc}. For YCSB, we use one table with 10 columns and 1,000,000 rows, and we configure it with 256 connections. To make it a transactional benchmark, we wrap operations within transactions and let each transaction contain 10 operations. We evaluate three different variations of YCSB workload: 1) YCSB-MC (medium contention): 80\% reads and 20\% writes with a hotspot of 10\% tuples that are accessed by $\sim$60\% of all queries ($\theta=0.8$ in Zipfian distribution). 2) YCSB-HC (high contention): 50\% reads and 50\% writes with a hotspot of 10\% tuples that are accessed by $\sim$75\% of all queries ($\theta=0.9$). 3) YCSB-RO (read only): all read queries and a uniform access distribution.
For the TPC-C benchmark, we configure it with 800 warehouses and 120 client connections for sending query requests. Since \calvin and \aria do not provide SQL engine (do not support interactive queries and complex queries, \eg join and range scan), they can only support New-Order transactions and Payment transactions. We follow \cite{lu2020aria} to make a TPC-C benchmark by mixing 50\% New-Order and 50\% Payment transactions. 
\calvinfs, \slog, and Anna do not support this TPC-C benchmark, so we do not run the TPC-C benchmark on them.

\begin{figure*}
\vspace{-0.2in}
		\centering
		\subfloat[YCSB-MC Throughput]{\label{fig:cross-region:throughput}
		\includegraphics[width=0.24\textwidth]{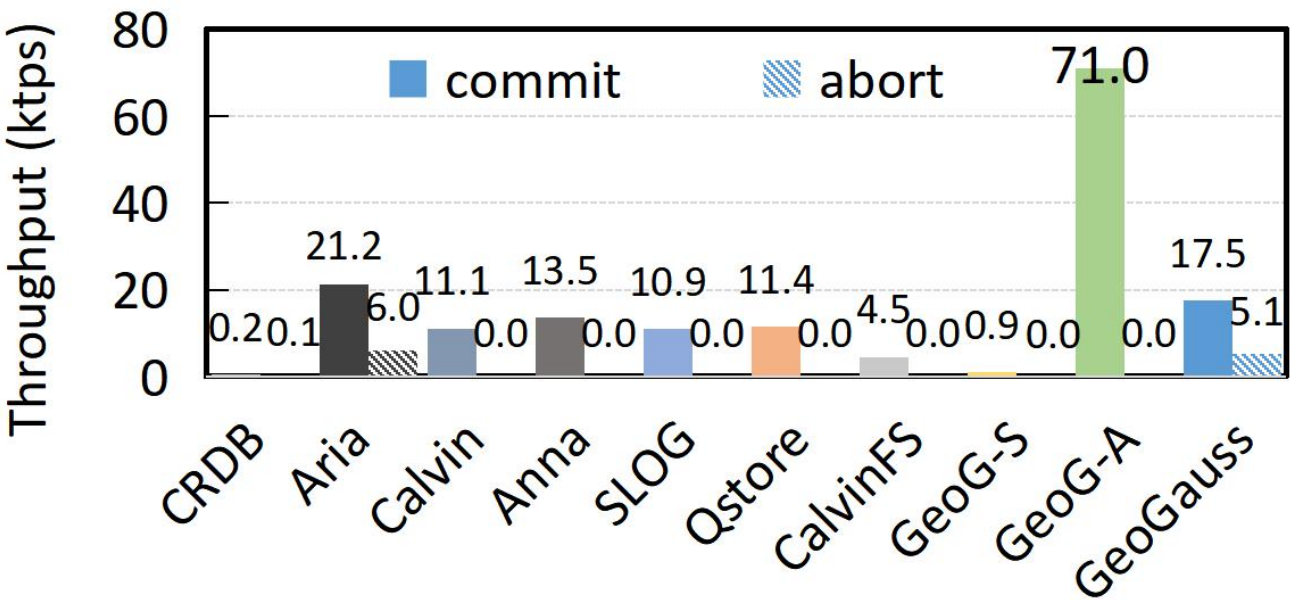}}
		\subfloat[YCSB-HC Throughput]{\label{fig:cross-region:throughput}
		\includegraphics[width=0.24\textwidth]{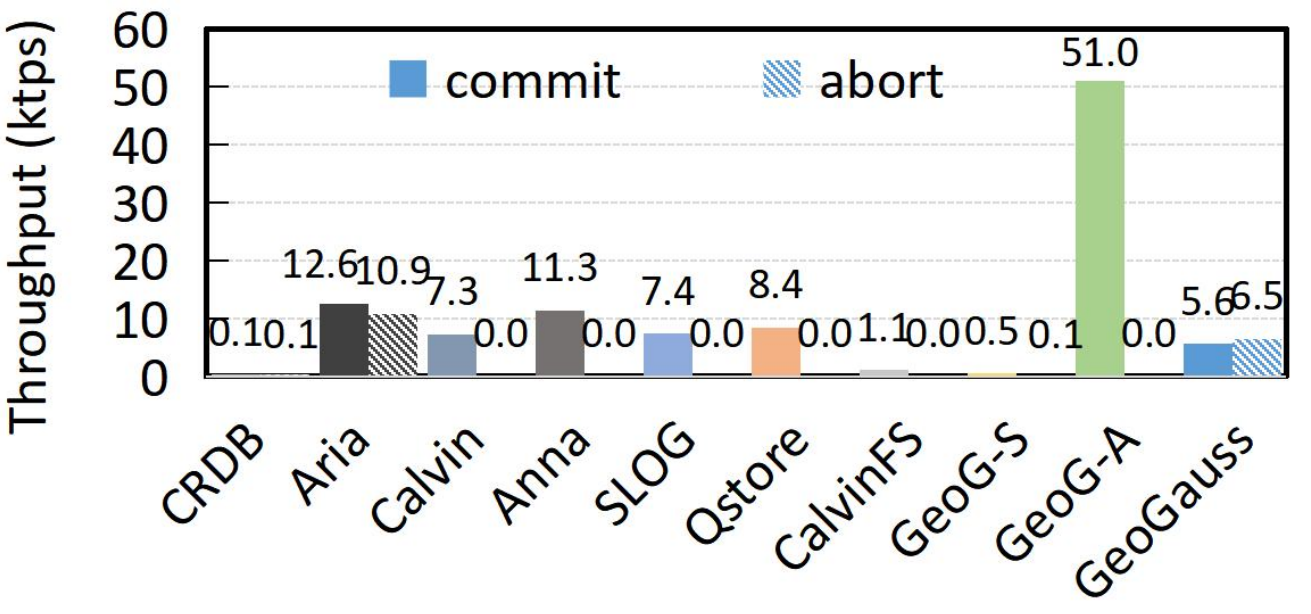}}
		\subfloat[YCSB-RO Throughput]{\label{fig:cross-region:throughput}
		\includegraphics[width=0.24\textwidth]{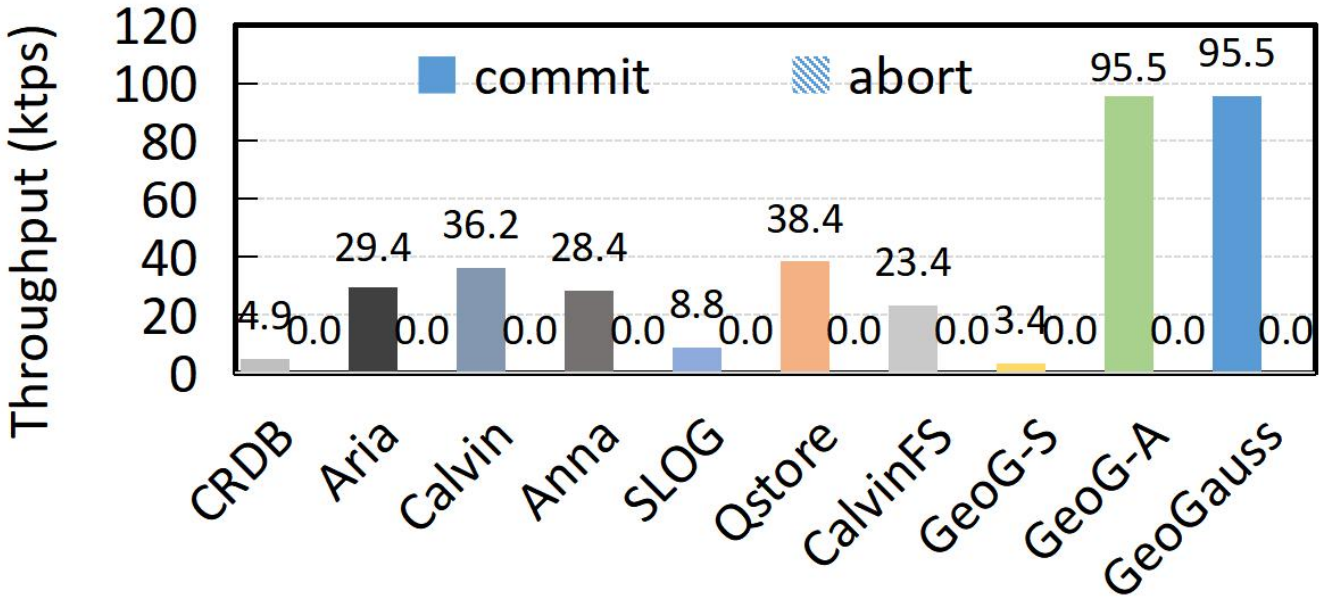}}
		\subfloat[TPC-C Throughput]{\label{fig:cross-region:throughput}
		\includegraphics[width=0.24\textwidth]{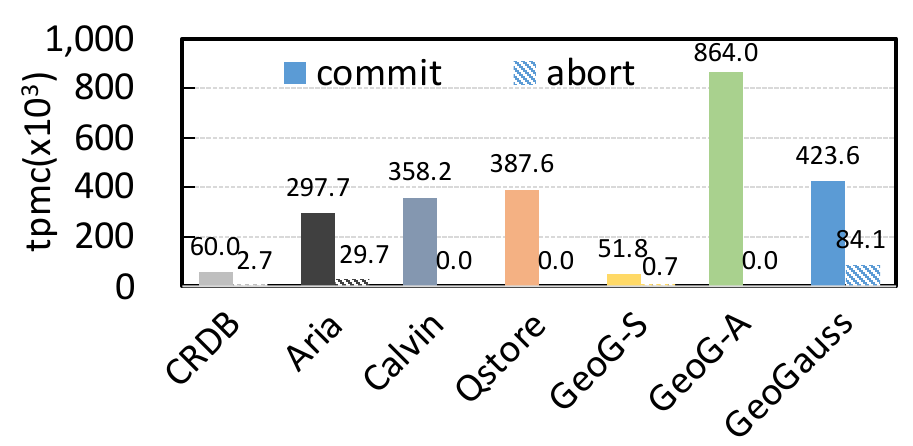}}
  \\\vspace{-0.1in}
		\subfloat[YCSB-MC Latency]{\label{fig:cross-region:latency}
		\includegraphics[width=0.24\textwidth]{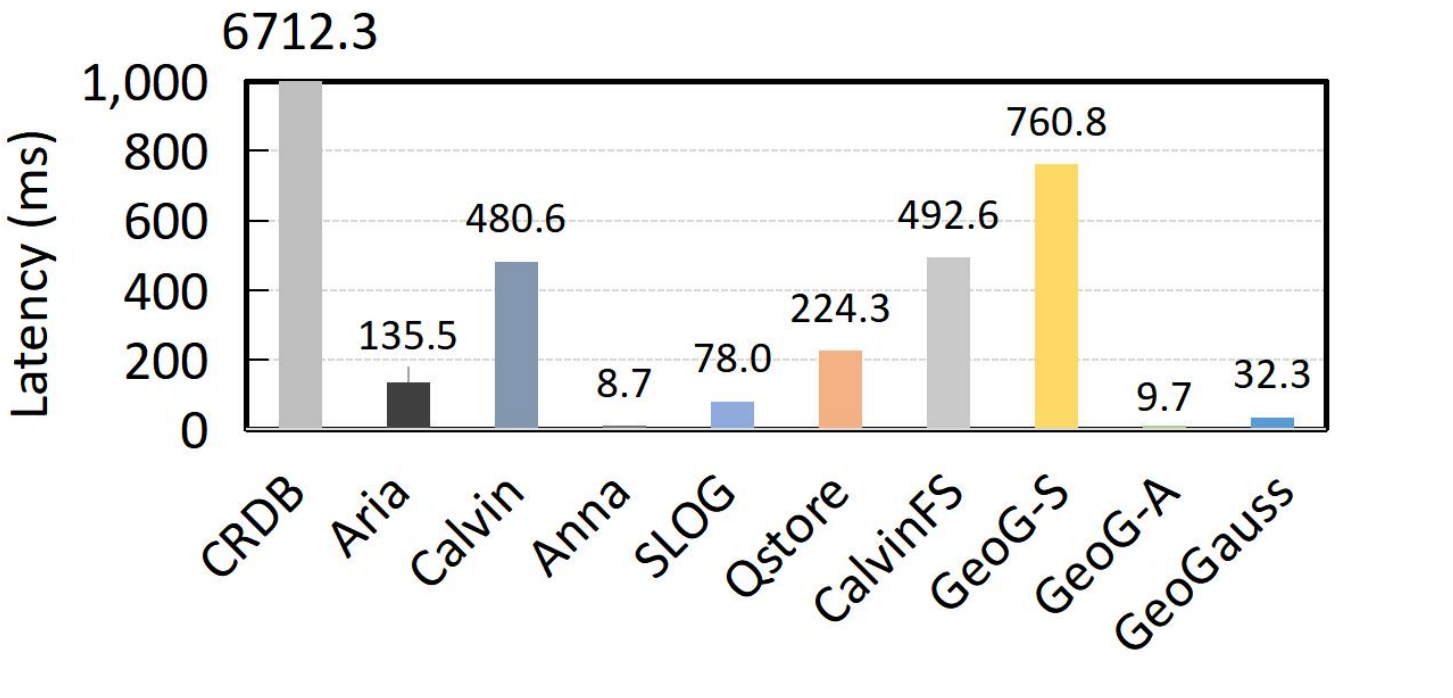}}
		\subfloat[YCSB-HC Latency]{\label{fig:cross-region:latency}
		\includegraphics[width=0.24\textwidth]{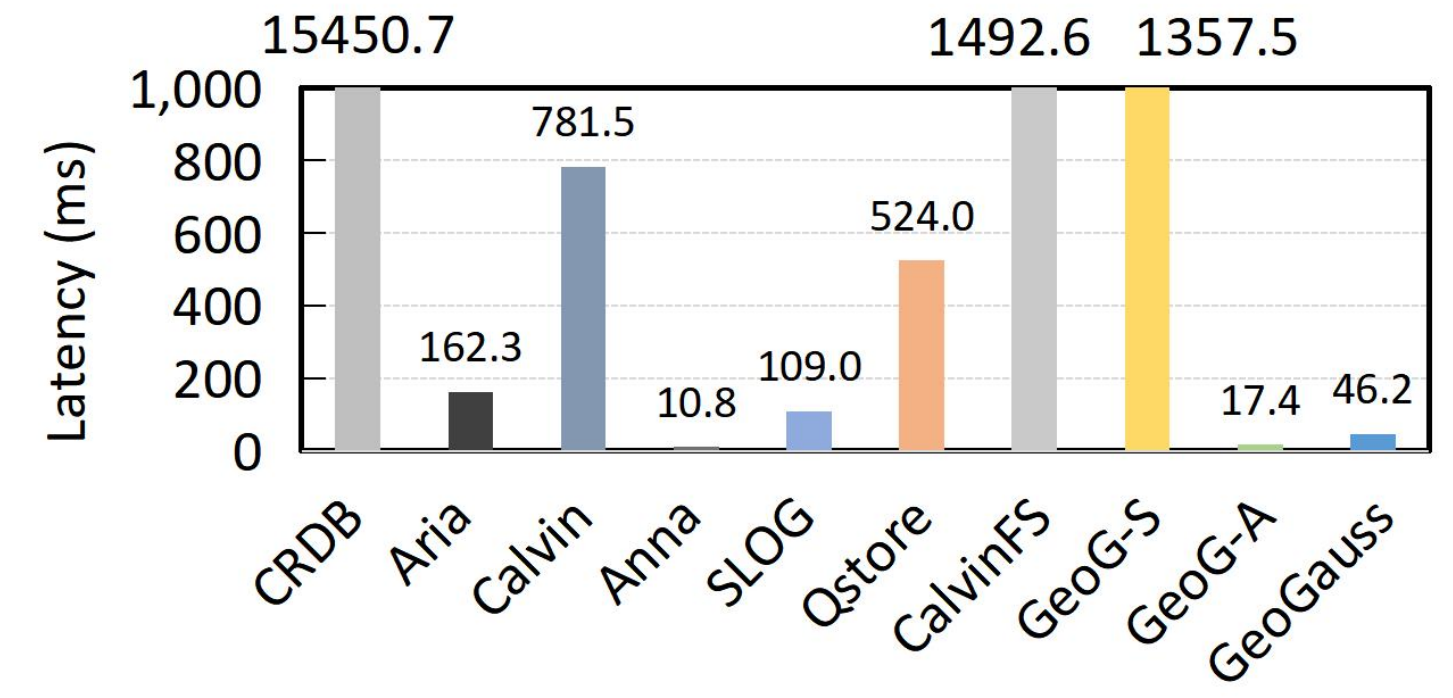}}
		\subfloat[YCSB-RO Latency]{\label{fig:cross-region:latency}
		\includegraphics[width=0.24\textwidth]{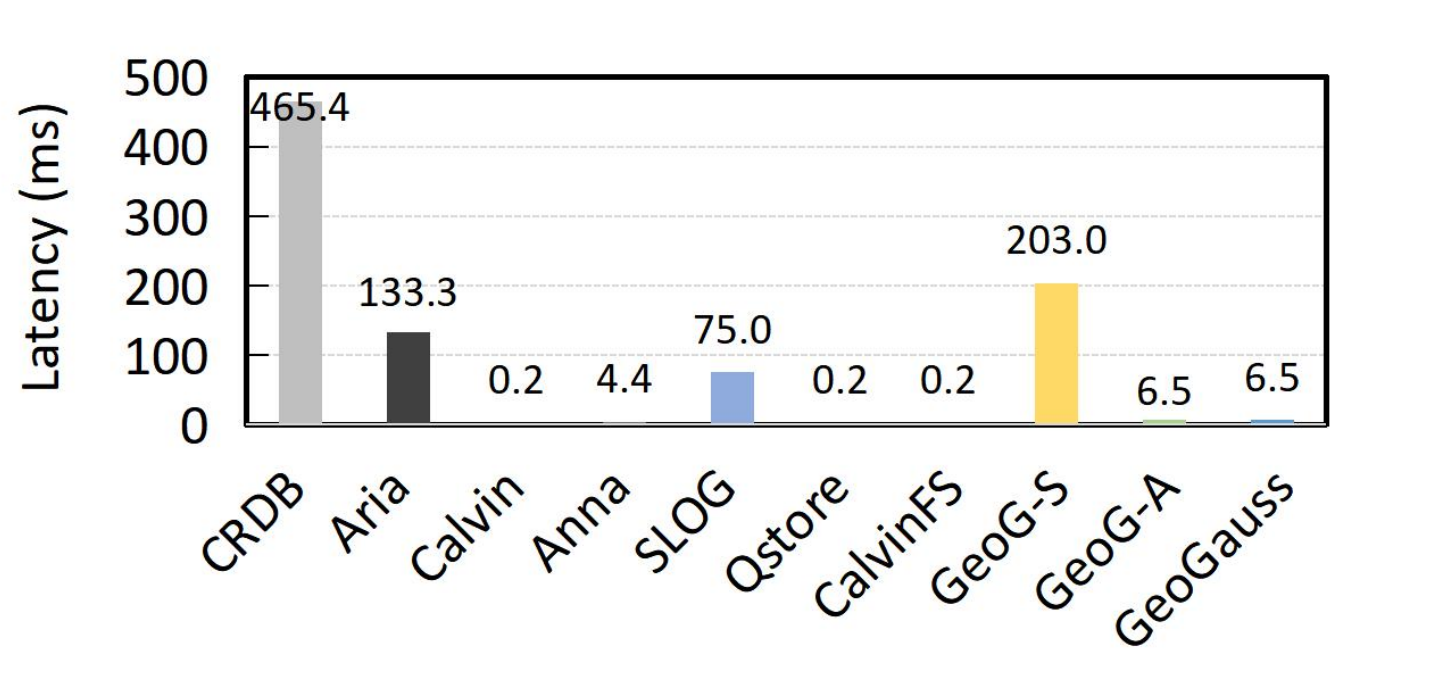}}
		\subfloat[TPC-C Latency]{\label{fig:cross-region:latency}
		\includegraphics[width=0.235\textwidth]{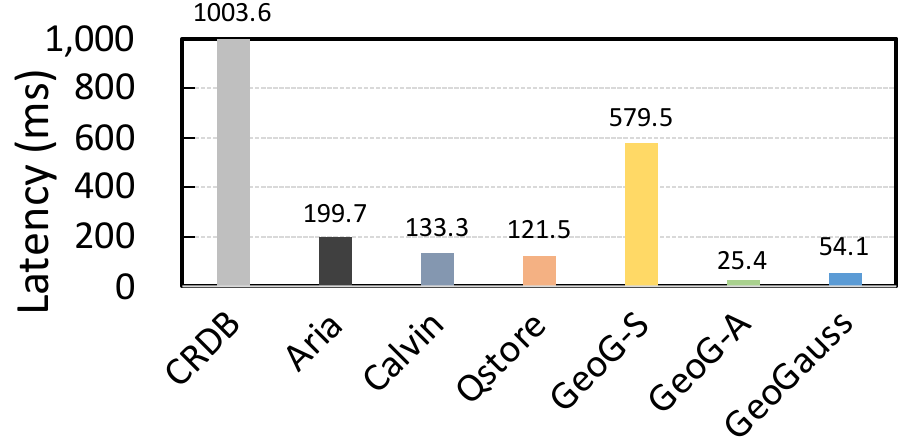}}
        \vspace{-0.1in}
		\caption{Comparison results on throughput and latency.}
		\label{fig:cross-region}
		\vspace{-0.15in}
\end{figure*}

\subsection{Cross-Region Comparison Results}
We run comparison experiments on our cross-region cluster. 
The throughput for successfully committed and aborted transactions and the average latency are reported in Figure \ref{fig:cross-region}. \oursys shows higher throughput over most of the other systems in the YCSB-MC workload, and achieves the highest throughput in read-intensive workloads. \oursysa is faster than \oursys, but it cannot guarantee strong consistency. Both \anna and \oursysa cannot respond to users with committed or aborted notifications since the state is not known with a time constraint (they have no aborted transactions). Strictly speaking, they are not transactional systems. \oursyss is a highly synchronized variant of \oursys, which guarantees strong consistency but at the expense of performance. \crdb is slow due to its expensive coordination cost under a geo-distributed environment. \calvin and \aria show much higher throughput than \crdb. \calvinfs, an extension of \calvin, uses Quorum protocol to achieve replica consistency, resulting in reduced performance compared with \calvin. \qstore reduces the scheduling overhead of \calvin with limited performance improvement because the main overhead in a geo-distributed environment is coordination rather than scheduling. \slog shows poorer performance than \calvin and \aria, because it requires to send cross-shard transactions to a single node for determining the global order, which hurts performance a lot under a cross-region scenario without a locality-aware sharding scheme. The writes and linearizable reads of a data item in \slog must be directed to its master replica, so the throughput on YCSB-RO is similar to that on YCSB-MC. 
\oursys and \aria exhibit higher abort rates due to their optimistic execution logic. \oursys leverages optimistic execution based on a stale snapshot and validates its effects before committing, while \aria relies on dependency graph analysis to resolve conflicts. These tend to have higher abort rates than pessimistic locking methods, \eg \calvin. Note that, deterministic databases and their extensions do not support the standard TPC-C benchmark. They are limited in interoperability and need to know what data will be accessed in advance.

Regarding the latency, \oursys exhibits superiority over the other strongly consistent systems on YCSB-MC, YCSB-HC, and TPC-C. \anna is a coordination-free KVS with eventual consistency. \oursysa only supports CRDT merge without epoch-based consistency, which only supports eventual consistency. Both of them do not need coordination, so their latency results are lower than the others. \crdb and \oursyss are with high latency due to the long waiting time under a geo-distributed environment.
\calvin and \aria need additional time for scheduling these batched transactions to achieve deterministic execution, so they result in higher latency than \oursys. 
\blue{On the YCSB-RO workload, \calvin and its extensions (\calvinfs and \qstore) show very low read latency because they directly return local read data and do not have the input parsing overhead. \calvin does not offer full SQL support, while the input query in \oursys needs to go through SQL parser and optimizer, so the latency in \oursys is higher than \calvin and its extensions. \aria requires a preprocessing step to perform dependency analysis for input queries, so \aria exhibits much higher latency than \calvin and its extensions.}

\begin{table}
\vspace{-0.2in}
    \caption{Runtime breakdown of a transaction (TPC-C).}
    \vspace{-0.1in}
    \label{tab:full_tpcc}
    \centering
    \small
    {\renewcommand{\arraystretch}{1.0}
    \begin{tabular}{l c c c  }
        \toprule
        {\textbf{}} &
        {\textbf{\oursyss}} &
        {\textbf{\oursysa}} &
        {\textbf{\oursys}} \\
        \midrule
SQL Parse  & 4.6 ms & 4.6 ms & 4.6 ms \\
Execute & 5.8 ms & 6.5 ms & 4.8 ms\\
Wait & \textbf{564.2 ms} & \textbf{0 ms} & \textbf{34.1 ms}  \\
Merge & 4.0 ms & 10.9 ms & 9.4 ms \\
Log & 0.8 ms & 6.5 ms & 4.7 ms \\
        \bottomrule
    \end{tabular}
    }
    \vspace{-0.05in}
\end{table}

\subsection{Performance Gain Analysis}
\label{sec:expr:long}
In our system, the transaction processing mainly contains five phases, including SQL parsing, transaction execution, waiting for the most recent snapshot being generated (Line 22 in Algorithm 1) or all remote/local transactions of the same \cen being applied (Line 24 in Algorithm 1), merging (Algorithm 2), and logging for duration. We study the cost of different phases to investigate the bottleneck. Table \ref{tab:full_tpcc} shows the average time spent in each phase of a successfully committed TPC-C transaction. We can see that the waiting phase dominates the runtime in \oursyss. By leveraging optimistic asynchronous execution (still needing synchronous validation), \oursys reduces the wait time dramatically. This is the key to improving the performance while at the same time guaranteeing strong consistency on a per-epoch basis (by synchronous validation). On the contrary, \oursysa does not wait for synchronization, so it does not provide strong consistency. The merge time and the logging time of \oursyss are shorter because the processed transactions in \oursyss are much less than the other two. 

\begin{figure}[h]
\vspace{-0.25in}
		\centering
		\subfloat[Number of commit transactions]{\label{fig:time:throughput}
		\includegraphics[width=0.4\textwidth]{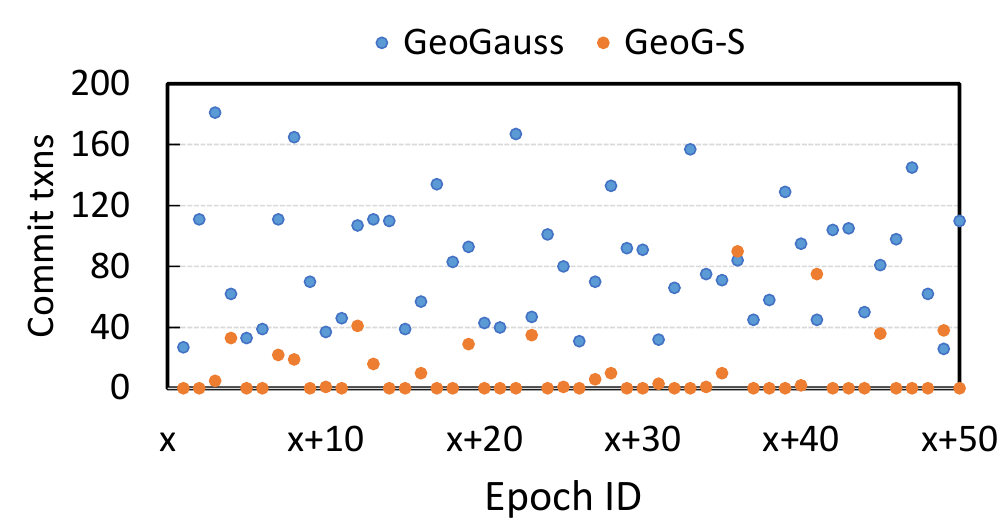}}
		\subfloat[Latency]{\label{fig:time:latency}
		\includegraphics[width=0.4\textwidth]{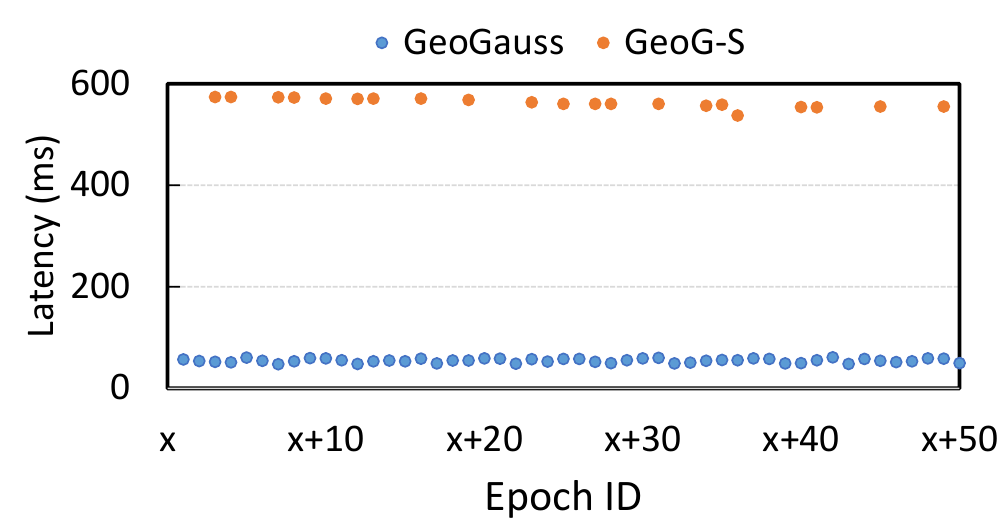}}
            \vspace{-0.1in}
		\caption{Throughput and latency of each epoch (TPC-C).}
		\label{fig:time_throughput_latency}
            \vspace{-0.15in}
\end{figure}

As analyzed above, the synchronous execution in \oursyss which requires processing the transactions of epoch $i$ based on the snapshot $(i-1)$ can greatly impact the performance. To investigate the impact of synchronous execution, we further measure the number of \red{committed} transactions and the average latency of each epoch in \oursys and \oursyss. Figure \ref{fig:time:throughput} shows the number of committed transactions in a number of consecutive epochs. In \oursyss, there is no commit transaction in many epochs and there are commit transactions every other 2-4 epochs. This is because the single-trip time is around 30 ms and \oursyss stops serving when waiting for the remote write sets of the previous epoch. While \oursys which exploits optimistic asynchronous execution can continuously commit transactions. Figure \ref{fig:time:latency} shows the average latency of the epochs that have \red{committed} transactions. The latency is much longer than the single-trip delay ($\sim$580 vs. $\sim$30 ms) due to the ripple effect.

\begin{figure}
\vspace{-0.3in}
		\centering
		\subfloat[Delay = 20ms]{\label{fig:long_transaction:ycsb_20ms}
		\includegraphics[width=0.4\textwidth]{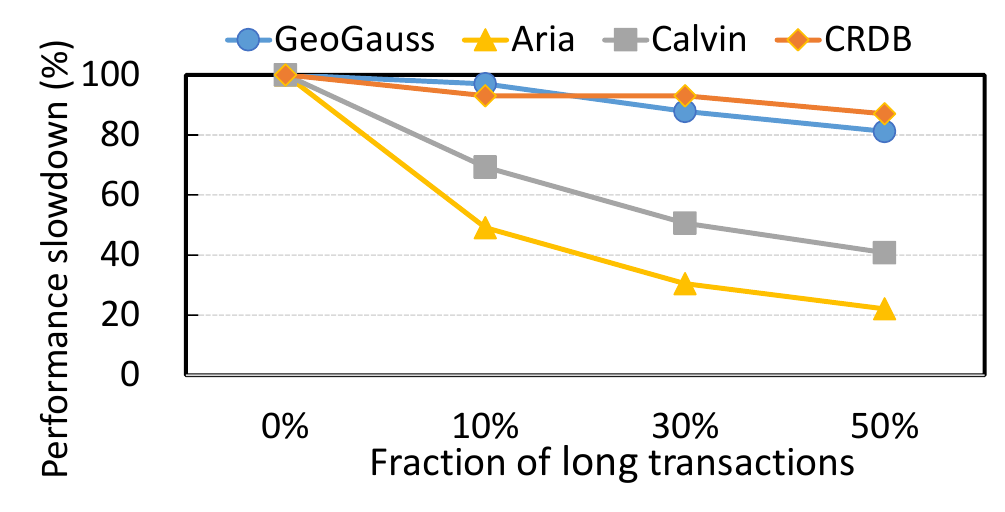}}
		\subfloat[Delay = 100ms]{\label{fig:long_transaction:ycsb_100ms}
		\includegraphics[width=0.4\textwidth]{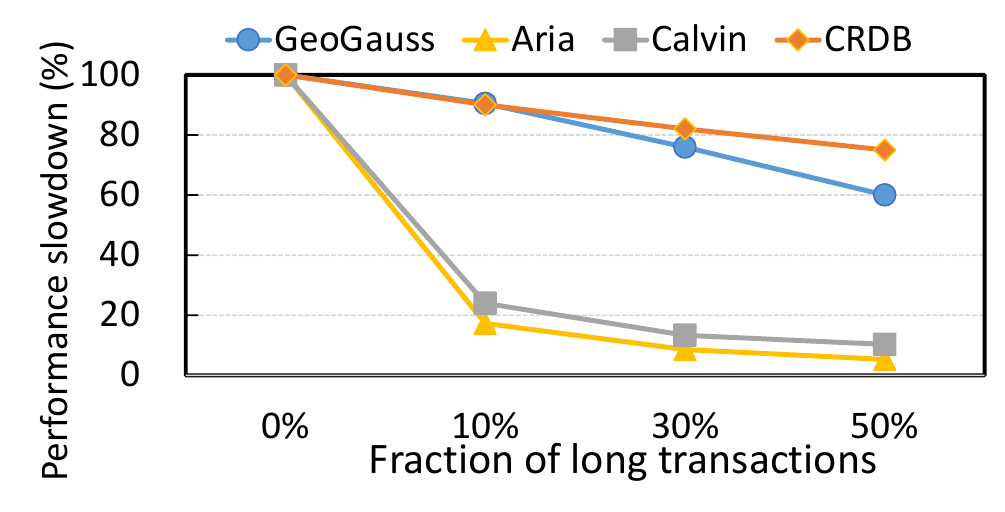}}
        \vspace{-0.1in}
		\caption{Effect of long transactions (YCSB-MC).}
		\label{fig:long_transaction}
		\vspace{-0.1in}
\end{figure}

\subsection{Long Running Transactions}
\label{sec:expr:long}
As discussed in the distinctions from deterministic databases (see Section \ref{sec:discuss}), \oursys should exhibit superiority over deterministic databases when processing long running transactions, since a barrier exists across batches in deterministic databases. In this experiment, we manually add a fixed 20 ms/100 ms delay to a randomly selected YCSB-MC transaction to simulate long running transactions. Figure \ref{fig:long_transaction} plots the system’s performance slowdown (with respect to throughput) when varying the portion of long transactions. The throughput of \aria and \calvin both decrease a lot when processing more long transactions. The performance slowdown of deterministic databases is even more significant when processing longer transactions as shown in Figure \ref{fig:long_transaction:ycsb_100ms}, say 80\% slowdown for 10\% long transactions. \oursys is more robust to long transactions. \blue{This is because that transactions in \oursys optimistically read data and write updates in their private cache, and new transactions are processed concurrently with the execution of long transactions.} It is worth mentioning that the synchronization point in our algorithm is for ensuring the completeness of remote updates but not a global barrier across batches of transactions. \crdb is robust to long transactions though its latency is much longer than ours, because the (manually set) delay is hidden in its \textit{parallel commit} phase.

\begin{table}[h]
    \vspace{-0.1in}
    \caption{Average WAN traffic per transaction (KB/txn)}
    \vspace{-0.1in}
    \label{tab:wan_traffic}
    \centering
    \small
    {\renewcommand{\arraystretch}{1.0}
    \begin{tabular}{l c c c c }
        \toprule
        {\textbf{}} &
        {\textbf{YCSB-RO}} &
        {\textbf{YCSB-MC}} &
        {\textbf{YCSB-HC}} &
        {\textbf{TPC-C}} \\
        \midrule
\blue{\oursys} & \blue{0} & \blue{0.28} & \blue{0.5} & \blue{0.6}  \\
\blue{\calvin} & \blue{0.13} & \blue{0.19} & \blue{0.28} & \blue{0.24} \\
        \bottomrule

    \end{tabular}
    }
\end{table}
\vspace{-0.35in}

\blue{\subsection{WAN Traffic}}
\label{sec:expr:WAN}
\blue{\oursys produces more WAN traffic than deterministic databases, because \oursys sends outputs (\ie write sets) instead of inputs (SQL statements). Table \ref{tab:wan_traffic} reports the average WAN traffic for each transaction in \oursys and in \calvin. The reported numbers are the average size of each transaction after compression by Gzip \cite{gzip} (see Section \ref{sec:impl:parallel}). As the write data size is bigger, transferring the write set results in more network traffic than transferring the input SQLs. However, we find that \calvin cannot fully utilize the WAN bandwidth (only about 25 Mbps on TPC-C, which is less than 100 Mbps WAN bandwidth). The bottleneck of deterministic databases is the scheduling for deterministic execution and the execution cost, especially for long transactions. In \oursys, we rely on CRDT for merging write sets to achieve determinism (regardless of arrival order and scheduling order), which is more efficient. Furthermore, we let each replica execute local SQLs first and only exchange outputs, so the execution cost is disseminated.}

\subsection{Varying Configurations}
\label{sec:5:threads}
\subsubsection{Epoch Length}
\label{sec:5:epoch}
As discussed in Section \ref{sec:impl:parallel}, \oursys uses one thread per transaction (\ie per connection) and will not release the connection until the transaction is committed or aborted. Thus, a longer epoch could lead to a phenomenon that all the connections are occupied by transactions waiting for confirmation and the system may stop serving for a period. On the other hand, it also affects the latency since a submitted write request must wait for the epoch snapshot to be generated before responding to users. A long epoch will result in long latency, while a short epoch may result in too frequent conflicts merging and also degrades performance.
\begin{figure}[h]
\vspace{-0.2in}
		\centering
		\subfloat[YCSB-MC]{\label{fig:epoch:ycsb}
		\includegraphics[width=0.4\textwidth]{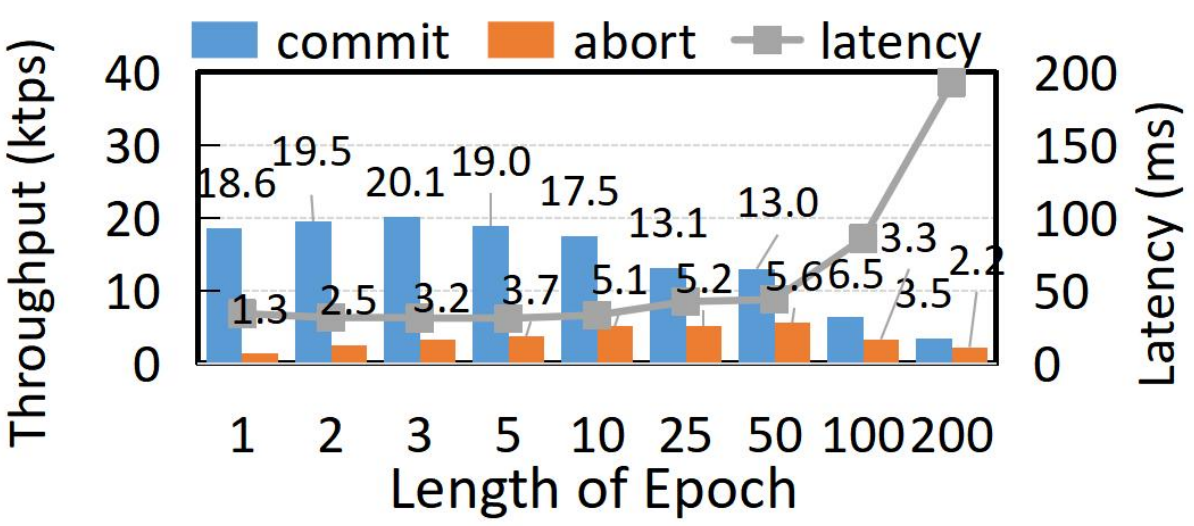}}
		\subfloat[TPC-C]{\label{fig:epoch:tpcc}
		\includegraphics[width=0.4\textwidth]{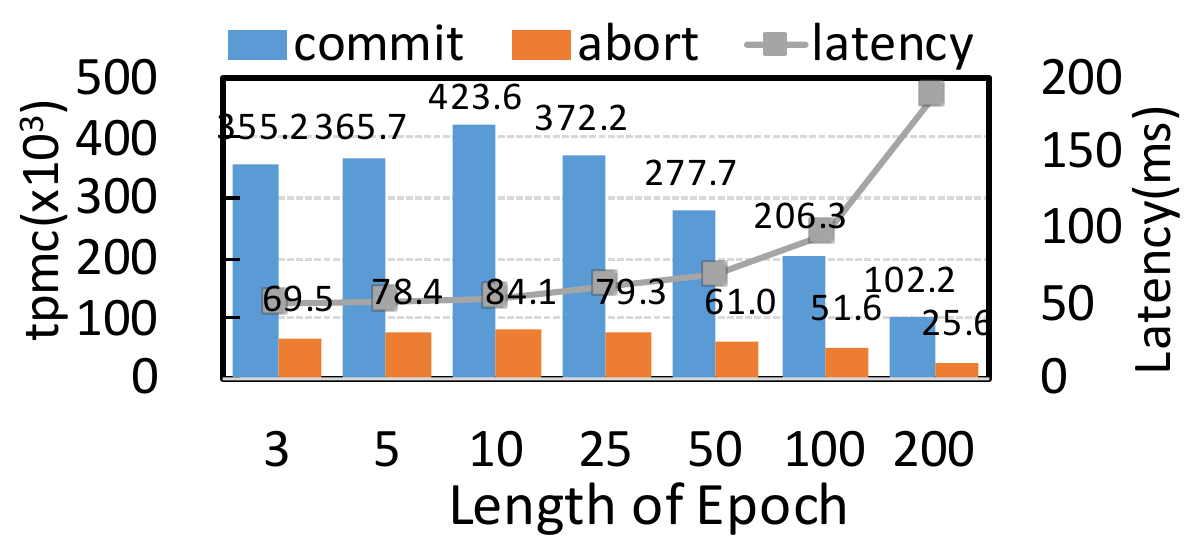}}
		\vspace{-0.15in}
		\caption{Effect of epoch length.}
		\label{fig:epoch}
		\vspace{-0.15in}
\end{figure}

Figure \ref{fig:epoch} shows the throughput and latency results of YCSB-MC and TPC-C workloads with different epoch lengths ranging from 1 ms to 200 ms. The throughput results are under expectation as analyzed above. The latency is determined on one hand by the single-trip delay when the epoch length is short; on the other hand by the epoch length setting when the epoch length is long.

\begin{figure}[h]
\vspace{-0.2in}
		\centering
		\subfloat[YCSB-MC]{\label{fig:isolation:ycsb}
		\includegraphics[width=0.4\textwidth]{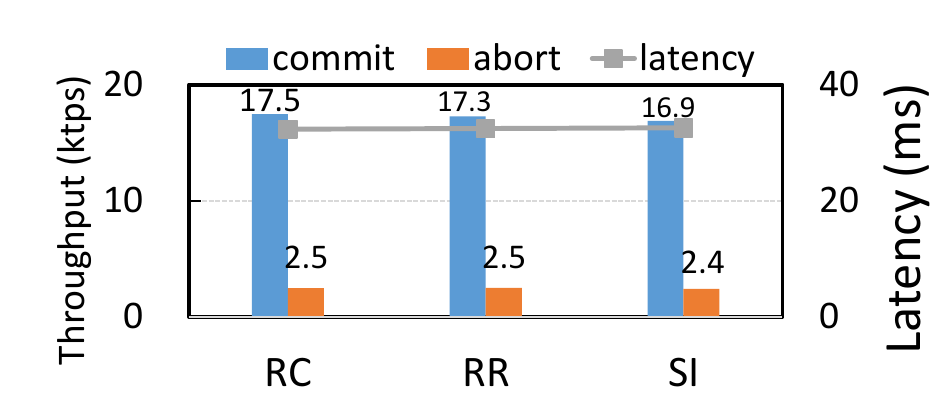}}
		\subfloat[TPC-C]{\label{fig:isolation:tpcc}
		\includegraphics[width=0.4\textwidth]{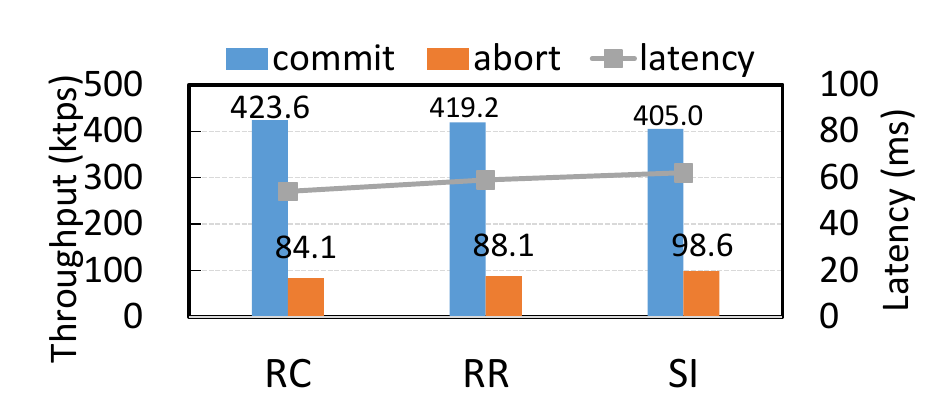}}
		\vspace{-0.1in}
		\caption{Performance with different isolation levels.}
		\label{fig:isolation}
		\vspace{-0.2in}
\end{figure}
\subsubsection{Isolation}
We study the performance when adopting different isolation levels in this experiment. Figure \ref{fig:isolation} shows the results under YCSB-MC and TPC-C workloads. There is not too much difference in the throughput and latency results under different isolation levels, except that the abort rate is higher with higher isolation levels. This is under expectation since \RR and \SI require a read set validation step that may increase the abort rate. 
\begin{figure}[h]
\vspace{-0.2in}
		\centering
		\subfloat[YCSB-MC]{\label{fig:real_contention:ycsb_mc}
		\includegraphics[width=0.4\textwidth]{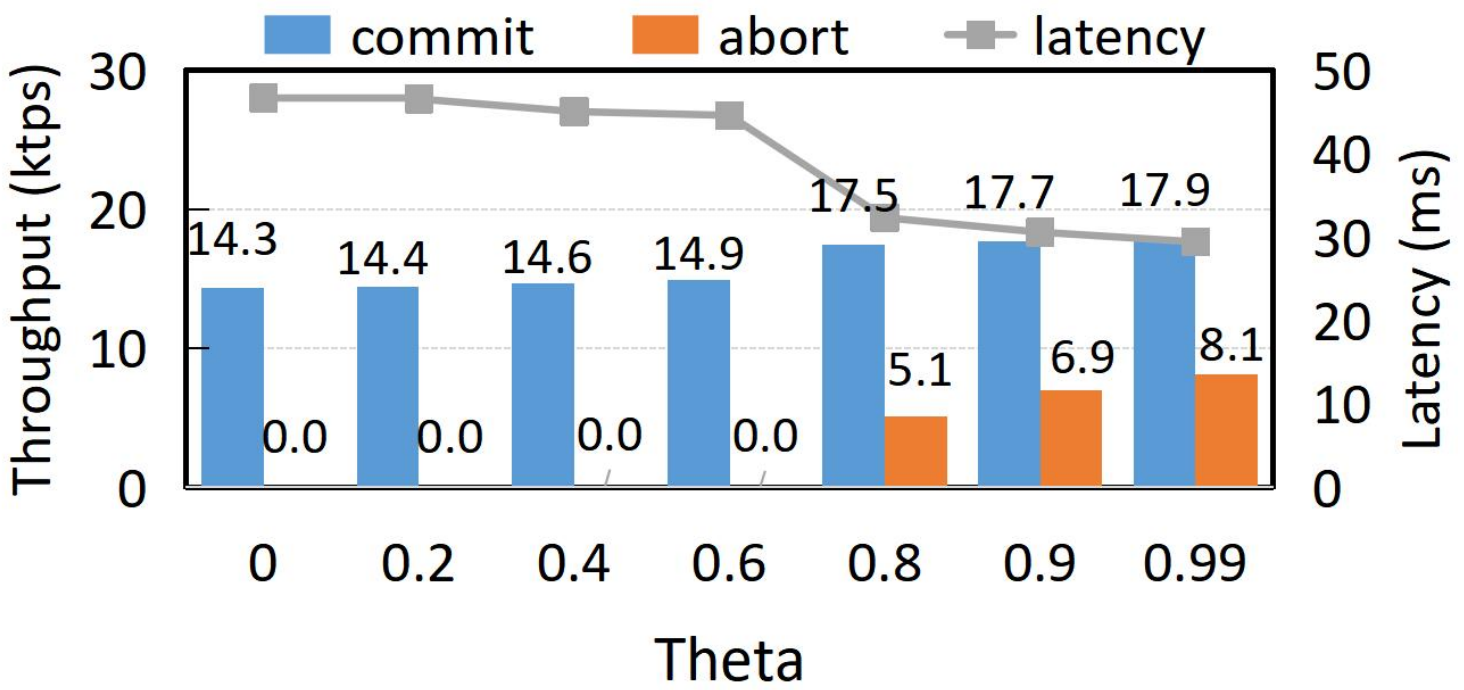}}
		\subfloat[YCSB-HC]{\label{fig:real_contention:ycsb_hc}
		\includegraphics[width=0.4\textwidth]{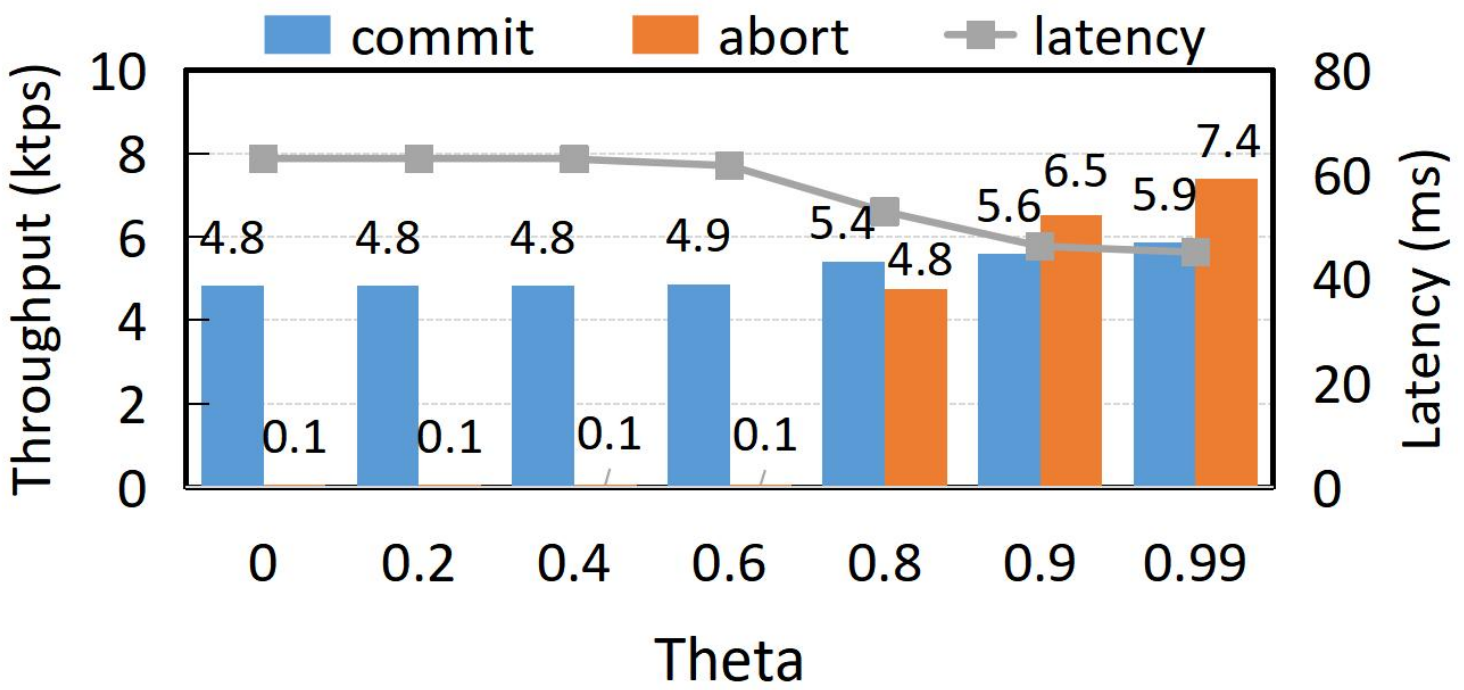}}
		\vspace{-0.1in}
  \caption{Performance with different contention levels.}
  \label{fig:contention}
  \vspace{-0.2in}
\end{figure}
\subsubsection{\blue{Contention Levels}}
To study the performance under different contention levels, we vary the $\theta$ parameter (the skew factor) from 0 to 0.99 in YCSB-MC and YCSB-HC workloads. As shown in Figure \ref{fig:contention}, the abort rate is higher when $\theta$ is bigger due to a higher probability of conflict, and the abort rate under the write-intensive workload YCSB-HC is even higher than in YCSB-MC. It is interesting that the throughput is a bit higher and the latency is a bit lower when $\theta$ is bigger. This is because that the aborted transactions resulted from conflicts will release the threads early without blocking (Line 24 in Algorithm \ref{alg:tx}), which allows serving more new transactions (that do not conflict with others) to commit. If the new transaction is read-only, it will be directly returned, which helps reduce the latency and improve the throughput.

\begin{figure}[t]
\vspace{-0.2in}
		\centering
		\subfloat[YCSB-MC (China)]{\label{fig:scalability:ycsb}
		\includegraphics[width=0.4\textwidth]{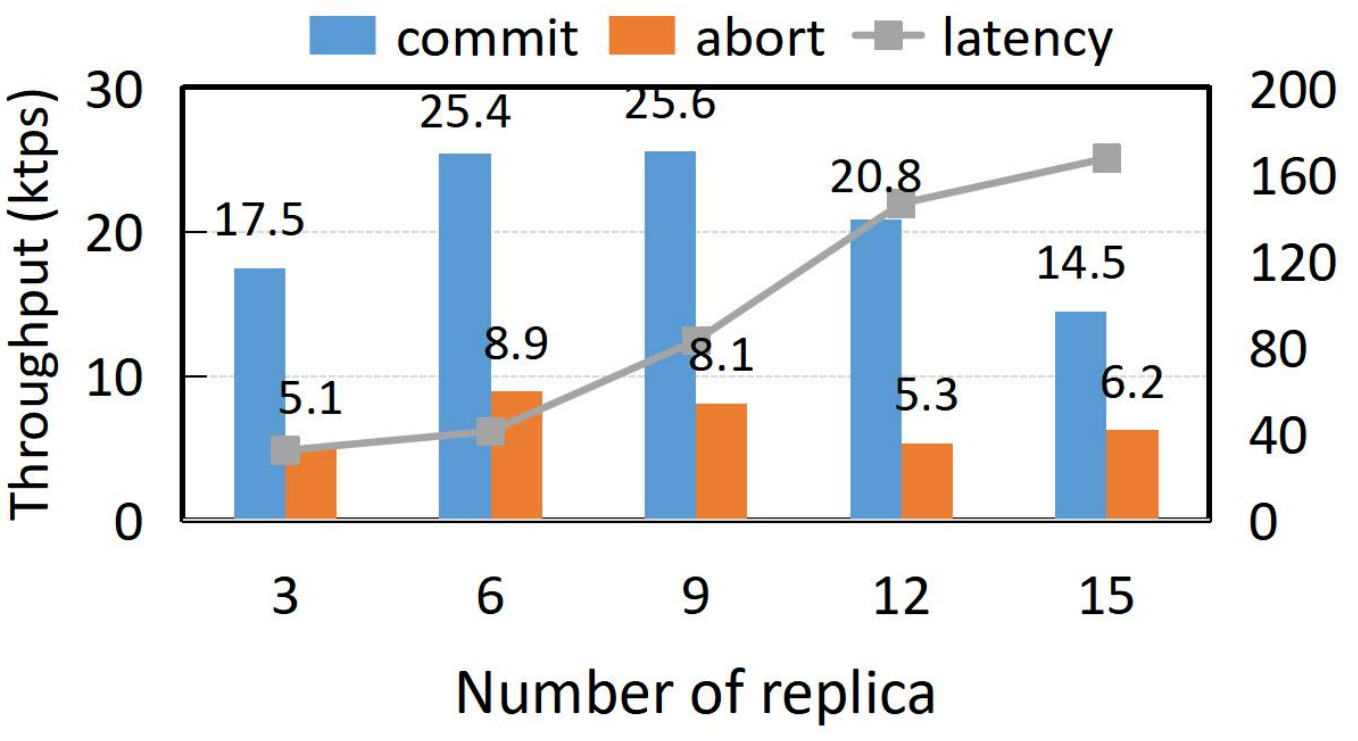}}
		\subfloat[YCSB-MC (WorldWide)]{\label{fig:scalability:continent}
            \includegraphics[width=0.4\textwidth]{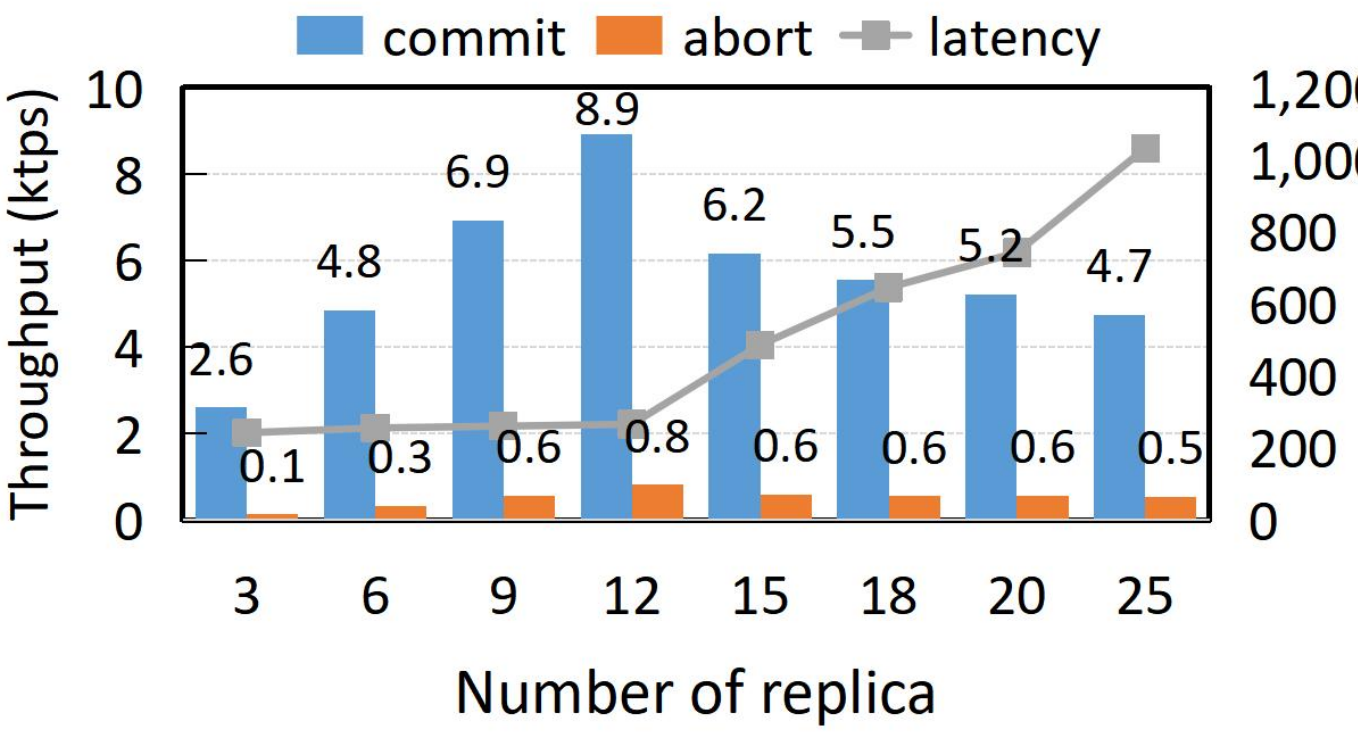}}
		\vspace{-0.1in}
		\caption{Scalability (YCSB-MC).}
		\label{fig:scalability}
		\vspace{-0.1in}
\end{figure}

\subsection{Scalability}
In this experiment, we study the scalability of \oursys. We scale the number of replica nodes from 3 to 15 and run YCSB-MC. As shown in Figure \ref{fig:scalability:ycsb}, the throughput is steadily increased when the number of replica nodes is no more than 9, then the throughput decreases. This is because more replicas will result in more replication overhead, which degrades the performance. the latency increases when using more replicas. This is because the epoch-based merge requires to receive updates from all replicas, which may lead to longer synchronization time. \blue{To investigate the scaling performance in a cross-continental setting. We set up a worldwide cluster that spreads across continents, including 25 nodes deployed in five data centers (London, Singapore, Tokyo, Silicon Valley, and Virginia). We scale the number of replica nodes from 3 to 25 and run YCSB-MC. The throughput and latency results are shown in Figure \ref{fig:scalability:continent}. We can see a similar trend with the scalability experiment in China. The peak throughput in the worldwide cluster is lower than in China, and the latency is much longer, which is under expectation.}

\begin{figure}[h]
\vspace{-0.15in}
    \subfloat[throughput]{\label{fig:throughput}
    \includegraphics[width=0.4\textwidth]{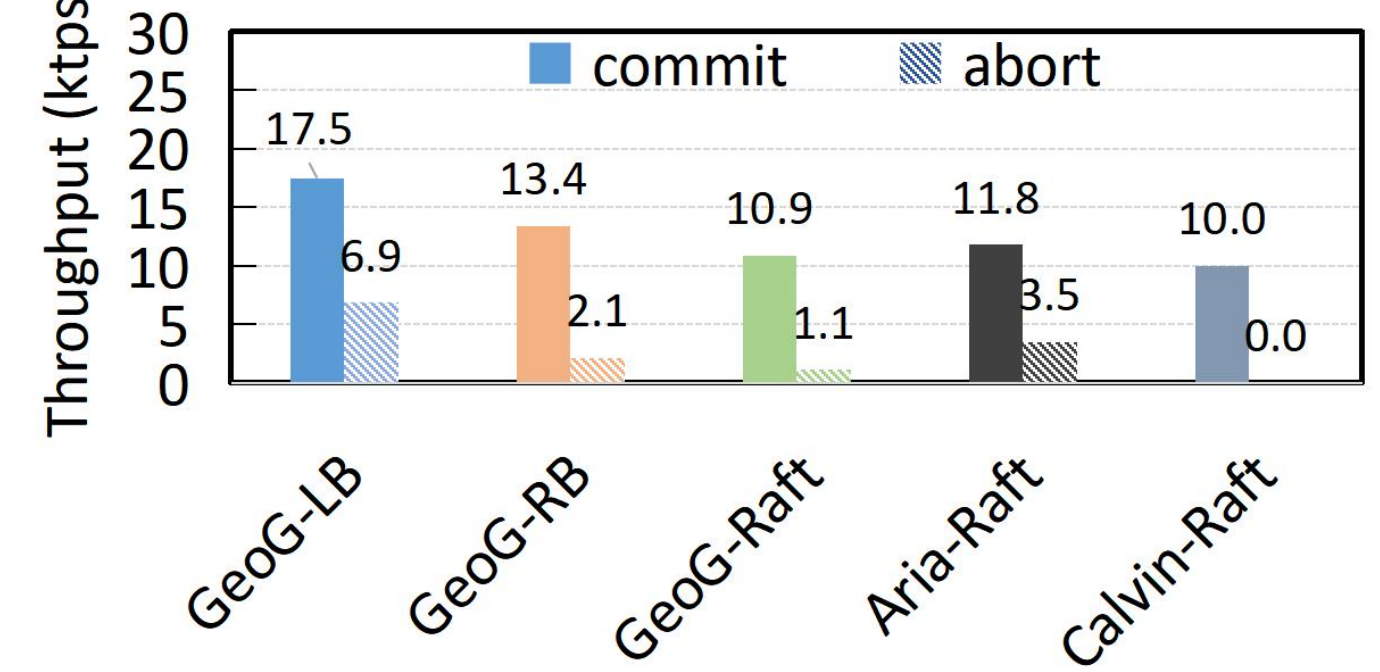}}
    \subfloat[latency]{\label{fig:throughput}
    \includegraphics[width=0.4\textwidth]{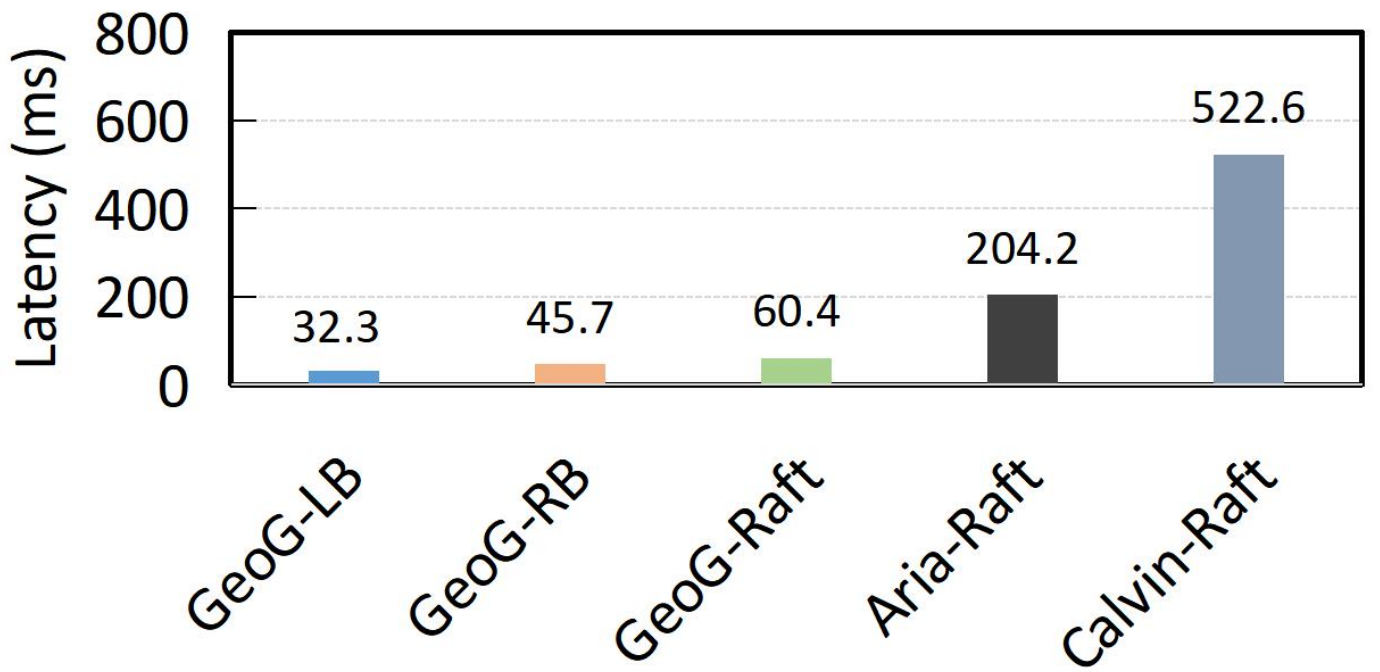}}
    \vspace{-0.1in}
    \caption{Performance with fault tolerance (YCSB-MC).}
    \label{fig:fault_tolerance_raft}
    \vspace{-0.1in}
\end{figure}

\subsection{\blue{Fault Tolerance}}
\label{sec:expr:fault}

In the above experiments, the fault tolerance supports of deterministic databases, \eg \calvin and \aria, are turned off for optimal performance, and \oursys only uses local write set backup server which does not affect performance. In this experiment, we turn on Raft replication in \calvin, \aria, and \oursys, and compare the performance of these systems under fault tolerance support (\ie \calvinha, \ariaha, \oursysha). We also measure the performance of \oursys with the other two weaker fault tolerance mechanisms, \ie local and remote write set backup server (\oursyslb and \oursysrb). As shown in \ref{fig:fault_tolerance_raft}, \oursysha shows comparable throughput with \calvinha and \ariaha. 
When adopting weaker fault tolerance mechanisms, \eg \oursyslb and \oursysrb, the throughput can be greatly increased. Regarding the latency results, \oursysha shows much lower latency than \calvinha and \ariaha owing to our optimistic execution scheme. As discussed in Section \ref{sec:imple:fault}, \oursyslb, \oursysrb, and \oursysha require $\thicksim$0.5 RTT, $\thicksim$1 RTT, and $\thicksim$1.5 RTT, respectively, so \oursyslb shows lower latency than \oursysrb and \oursysha.

\begin{figure}[h]
\vspace{-0.1in}
		\centering
		\subfloat[client-throughput]{\label{fig:client-throughput}
		\includegraphics[width=0.4\textwidth]{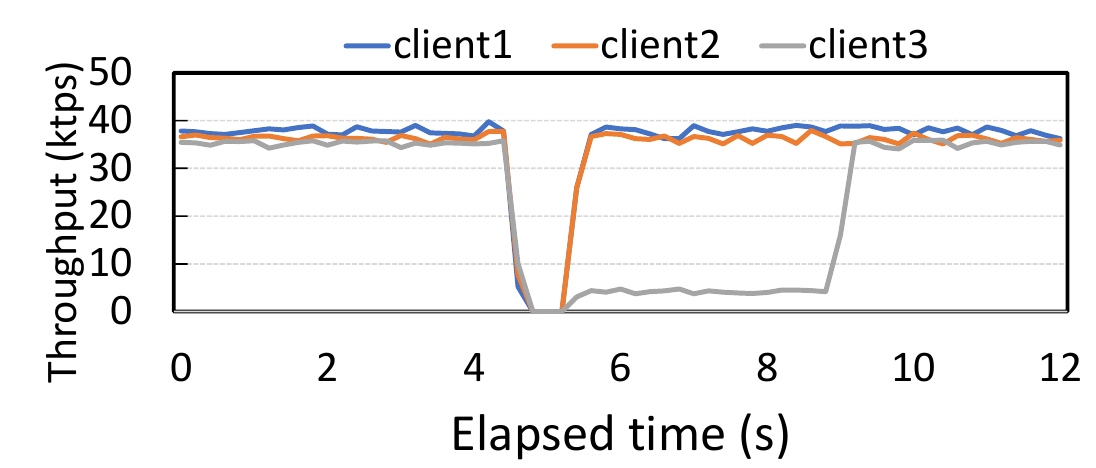}}
		\subfloat[client-latency]{\label{fig:client-lattency}
		\includegraphics[width=0.4\textwidth]{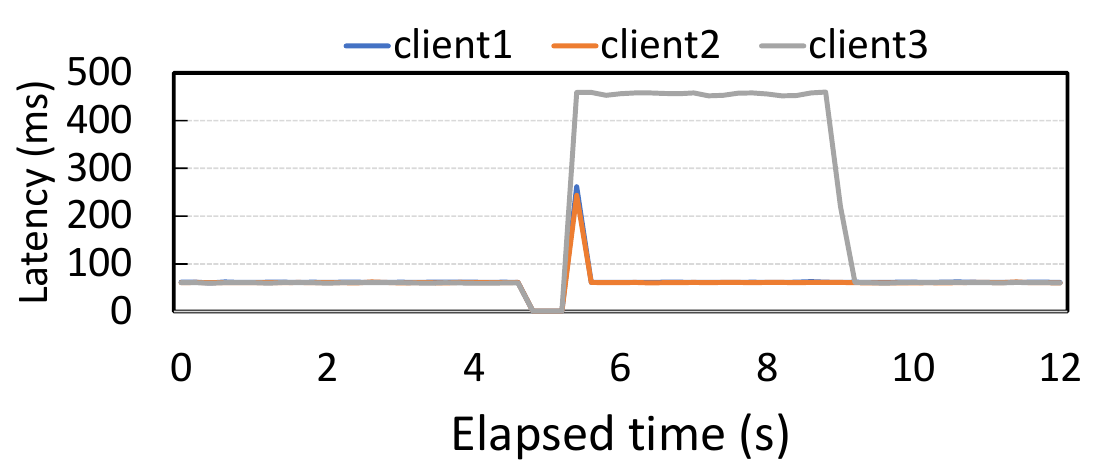}}
		\vspace{-0.1in}
		\caption{Performance under failures (YCSB-MC).}
		\label{fig:client}
		\vspace{-0.1in}
\end{figure}

\blue{To investigate the performance fluctuation under failures, we manually shut down a node and see how \oursys acts after failure. Figure \ref{fig:client} shows the changes in throughput and latency from each client's perspective, where we place a client that connects to the server node in each region. The temporary performance degradation during a node failure is due to the blocking of service, \ie waiting for the updates from the failed node. \oursys can quickly respond to this failure owing to our Raft-based membership management (with 500 ms timeout setup). The requests of client3 that were previously connected to the crashed node are routed to the nodes that are still providing services in other regions. But the transaction requests are executed in remote regions, resulting in decreased throughput and increased latency on client3. After the crashed node resumes, the Raft-based membership management will notice and let client3 connect to the node in the same region.}

%% file: related.tex
\section{Conclusion}
\label{sec:conclusion}

This paper presents \oursys, a strongly consistent and \blue{light-coordinated} geo-distributed transactional database with multi-master replication architecture. We have shown that the performance of geo-distributed transaction processing can be greatly improved by \oursys with strong consistency and weak isolation. By employing epoch-based 
output replication and optimistic asynchronous execution, our multi-master OCC algorithm can efficiently merge the conflicts and at the same time enforces strong consistency of replicas at the granularity of epochs. \oursys also overcomes the disadvantage of deterministic databases and supports interactive SQL execution, which has wider applicability. 





%% file: main.bbl

\begin{thebibliography}{74}


\ifx \showCODEN    \undefined \def \showCODEN     #1{\unskip}     \fi
\ifx \showDOI      \undefined \def \showDOI       #1{#1}\fi
\ifx \showISBNx    \undefined \def \showISBNx     #1{\unskip}     \fi
\ifx \showISBNxiii \undefined \def \showISBNxiii  #1{\unskip}     \fi
\ifx \showISSN     \undefined \def \showISSN      #1{\unskip}     \fi
\ifx \showLCCN     \undefined \def \showLCCN      #1{\unskip}     \fi
\ifx \shownote     \undefined \def \shownote      #1{#1}          \fi
\ifx \showarticletitle \undefined \def \showarticletitle #1{#1}   \fi
\ifx \showURL      \undefined \def \showURL       {\relax}        \fi
\providecommand\bibfield[2]{#2}
\providecommand\bibinfo[2]{#2}
\providecommand\natexlab[1]{#1}
\providecommand\showeprint[2][]{arXiv:#2}

\bibitem[cou(2022)]%
        {couchdb}
 \bibinfo{year}{2022}\natexlab{}.
\newblock \bibinfo{title}{Apache CouchDB}.
\newblock
\newblock
\urldef\tempurl%
\url{http://couchdb.apache.org/}
\showURL{%
\tempurl}


\bibitem[hba(2022)]%
        {hbase}
 \bibinfo{year}{2022}\natexlab{}.
\newblock \bibinfo{title}{Apache HBase}.
\newblock
\newblock
\urldef\tempurl%
\url{https://hbase.apache.org/}
\showURL{%
\tempurl}


\bibitem[ara(2022)]%
        {arangodb}
 \bibinfo{year}{2022}\natexlab{}.
\newblock \bibinfo{title}{ArangoDB}.
\newblock
\newblock
\urldef\tempurl%
\url{https://www.arangodb.com/}
\showURL{%
\tempurl}


\bibitem[ari(2022)]%
        {ariacode}
 \bibinfo{year}{2022}\natexlab{}.
\newblock \bibinfo{title}{Aria: A Fast and Practical Deterministic OLTP
  Database}.
\newblock
\newblock
\urldef\tempurl%
\url{https://github.com/luyi0619/aria}
\showURL{%
\tempurl}


\bibitem[bra(2022)]%
        {braft}
 \bibinfo{year}{2022}\natexlab{}.
\newblock \bibinfo{title}{Baidu braft}.
\newblock
\newblock
\urldef\tempurl%
\url{https://github.com/baidu/braft}
\showURL{%
\tempurl}


\bibitem[Cal(2022)]%
        {CalvinFScode}
 \bibinfo{year}{2022}\natexlab{}.
\newblock \bibinfo{title}{CalvinFS}.
\newblock
\newblock
\urldef\tempurl%
\url{http://https://github.com/kunrenyale/CalvinFS}
\showURL{%
\tempurl}


\bibitem[clo(2022)]%
        {cloudant}
 \bibinfo{year}{2022}\natexlab{}.
\newblock \bibinfo{title}{Cloudant}.
\newblock
\newblock
\urldef\tempurl%
\url{https://www.ibm.com/hk-en/cloud/cloudant}
\showURL{%
\tempurl}


\bibitem[ext(2022)]%
        {extremedb}
 \bibinfo{year}{2022}\natexlab{}.
\newblock \bibinfo{title}{ExtremeDB: Cluster Distributed Database System}.
\newblock
\newblock
\urldef\tempurl%
\url{https://www.mcobject.com/cluster/}
\showURL{%
\tempurl}


\bibitem[fau(2022)]%
        {fauna}
 \bibinfo{year}{2022}\natexlab{}.
\newblock \bibinfo{title}{FaunaDB}.
\newblock
\newblock
\urldef\tempurl%
\url{https://fauna.com/}
\showURL{%
\tempurl}


\bibitem[gal(2022)]%
        {galera}
 \bibinfo{year}{2022}\natexlab{}.
\newblock \bibinfo{title}{Galera Cluster for MySQL}.
\newblock
\newblock
\urldef\tempurl%
\url{https://galeracluster.com/}
\showURL{%
\tempurl}


\bibitem[gzi(2022)]%
        {gzip}
 \bibinfo{year}{2022}\natexlab{}.
\newblock \bibinfo{title}{GNU Gzip}.
\newblock
\newblock
\urldef\tempurl%
\url{https://www.gnu.org/software/gzip/}
\showURL{%
\tempurl}


\bibitem[gRP(2022)]%
        {gRPC}
 \bibinfo{year}{2022}\natexlab{}.
\newblock \bibinfo{title}{gRPC: A high performance, open source universal RPC
  framework}.
\newblock
\newblock
\urldef\tempurl%
\url{https://grpc.io/}
\showURL{%
\tempurl}


\bibitem[tun(2022)]%
        {tungsten}
 \bibinfo{year}{2022}\natexlab{}.
\newblock \bibinfo{title}{MySQL Tungsten}.
\newblock
\newblock
\urldef\tempurl%
\url{https://www.continuent.com/products/tungsten-replicator}
\showURL{%
\tempurl}


\bibitem[mys(2022)]%
        {mysql}
 \bibinfo{year}{2022}\natexlab{}.
\newblock \bibinfo{title}{MySQL's primary-secondary replication}.
\newblock
\newblock
\urldef\tempurl%
\url{https://dev.mysql.com/}
\showURL{%
\tempurl}


\bibitem[ope(2022)]%
        {opengauss}
 \bibinfo{year}{2022}\natexlab{}.
\newblock \bibinfo{title}{openGauss}.
\newblock
\newblock
\urldef\tempurl%
\url{https://opengauss.org/}
\showURL{%
\tempurl}


\bibitem[pgb(2022)]%
        {pgbdr}
 \bibinfo{year}{2022}\natexlab{}.
\newblock \bibinfo{title}{PostgreSQL BDR}.
\newblock
\newblock
\urldef\tempurl%
\url{https://wiki.postgresql.org/wiki/BDR_Project}
\showURL{%
\tempurl}


\bibitem[pro(2022)]%
        {protobuf}
 \bibinfo{year}{2022}\natexlab{}.
\newblock \bibinfo{title}{Protocol Buffers}.
\newblock
\newblock
\urldef\tempurl%
\url{https://developers.google.com/protocol-buffers}
\showURL{%
\tempurl}


\bibitem[red(2022)]%
        {redis}
 \bibinfo{year}{2022}\natexlab{}.
\newblock \bibinfo{title}{Redis CRDT}.
\newblock
\newblock
\urldef\tempurl%
\url{https://redis.com/blog/diving-into-crdts/}
\showURL{%
\tempurl}


\bibitem[ria(2022)]%
        {riak}
 \bibinfo{year}{2022}\natexlab{}.
\newblock \bibinfo{title}{Riak: Enterprise NoSQL Database}.
\newblock
\newblock
\urldef\tempurl%
\url{https://riak.com/}
\showURL{%
\tempurl}


\bibitem[sem(2022)]%
        {semisync}
 \bibinfo{year}{2022}\natexlab{}.
\newblock \bibinfo{title}{Semi-synchronous replication at facebook}.
\newblock
\newblock
\urldef\tempurl%
\url{http://yoshinorimatsunobu.blogspot.com/}
\showURL{%
\tempurl}


\bibitem[tpc(2022)]%
        {tpcc}
 \bibinfo{year}{2022}\natexlab{}.
\newblock \bibinfo{title}{TPC-C Homepage}.
\newblock
\newblock
\urldef\tempurl%
\url{https://www.tpc.org/tpcc/}
\showURL{%
\tempurl}


\bibitem[yug(2022)]%
        {yugabytedb}
 \bibinfo{year}{2022}\natexlab{}.
\newblock \bibinfo{title}{YugabyteDB: Distributed SQL Database}.
\newblock
\newblock
\urldef\tempurl%
\url{https://www.yugabyte.com/}
\showURL{%
\tempurl}


\bibitem[zer(2022)]%
        {zeromq}
 \bibinfo{year}{2022}\natexlab{}.
\newblock \bibinfo{title}{ZeroMQ: An open-source universal messaging library}.
\newblock
\newblock
\urldef\tempurl%
\url{https://zeromq.org/}
\showURL{%
\tempurl}


\bibitem[Abadi and Faleiro(2018)]%
        {abadi2018overview}
\bibfield{author}{\bibinfo{person}{Daniel~J Abadi} {and}
  \bibinfo{person}{Jose~M Faleiro}.} \bibinfo{year}{2018}\natexlab{}.
\newblock \showarticletitle{An overview of deterministic database systems}.
\newblock \bibinfo{journal}{\emph{Commun. ACM}} \bibinfo{volume}{61},
  \bibinfo{number}{9} (\bibinfo{year}{2018}), \bibinfo{pages}{78--88}.
\newblock


\bibitem[Abebe et~al\mbox{.}(2020a)]%
        {abebe2020dynamast}
\bibfield{author}{\bibinfo{person}{Michael Abebe}, \bibinfo{person}{Brad
  Glasbergen}, {and} \bibinfo{person}{Khuzaima Daudjee}.}
  \bibinfo{year}{2020}\natexlab{a}.
\newblock \showarticletitle{DynaMast: Adaptive dynamic mastering for replicated
  systems}. In \bibinfo{booktitle}{\emph{2020 IEEE 36th International
  Conference on Data Engineering (ICDE)}}. IEEE, \bibinfo{pages}{1381--1392}.
\newblock


\bibitem[Abebe et~al\mbox{.}(2020b)]%
        {abebe2020morphosys}
\bibfield{author}{\bibinfo{person}{Michael Abebe}, \bibinfo{person}{Brad
  Glasbergen}, {and} \bibinfo{person}{Khuzaima Daudjee}.}
  \bibinfo{year}{2020}\natexlab{b}.
\newblock \showarticletitle{MorphoSys: automatic physical design metamorphosis
  for distributed database systems}.
\newblock \bibinfo{journal}{\emph{Proceedings of the VLDB Endowment}}
  \bibinfo{volume}{13}, \bibinfo{number}{13} (\bibinfo{year}{2020}),
  \bibinfo{pages}{3573--3587}.
\newblock


\bibitem[Almeida et~al\mbox{.}(2015)]%
        {almeida2015efficient}
\bibfield{author}{\bibinfo{person}{Paulo~S{\'e}rgio Almeida},
  \bibinfo{person}{Ali Shoker}, {and} \bibinfo{person}{Carlos Baquero}.}
  \bibinfo{year}{2015}\natexlab{}.
\newblock \showarticletitle{Efficient state-based crdts by delta-mutation}. In
  \bibinfo{booktitle}{\emph{International Conference on Networked Systems}}.
  Springer, \bibinfo{pages}{62--76}.
\newblock


\bibitem[Almeida et~al\mbox{.}(2018)]%
        {almeida2018delta}
\bibfield{author}{\bibinfo{person}{Paulo~S{\'e}rgio Almeida},
  \bibinfo{person}{Ali Shoker}, {and} \bibinfo{person}{Carlos Baquero}.}
  \bibinfo{year}{2018}\natexlab{}.
\newblock \showarticletitle{Delta state replicated data types}.
\newblock \bibinfo{journal}{\emph{J. Parallel and Distrib. Comput.}}
  \bibinfo{volume}{111} (\bibinfo{year}{2018}), \bibinfo{pages}{162--173}.
\newblock


\bibitem[Alvaro et~al\mbox{.}(2017)]%
        {10.1145/3110214}
\bibfield{author}{\bibinfo{person}{Peter Alvaro}, \bibinfo{person}{Neil
  Conway}, \bibinfo{person}{Joseph~M. Hellerstein}, {and}
  \bibinfo{person}{David Maier}.} \bibinfo{year}{2017}\natexlab{}.
\newblock \showarticletitle{Blazes: Coordination Analysis and Placement for
  Distributed Programs}.
\newblock \bibinfo{journal}{\emph{ACM Trans. Database Syst.}}
  \bibinfo{volume}{42}, \bibinfo{number}{4}, Article \bibinfo{articleno}{23}
  (\bibinfo{date}{oct} \bibinfo{year}{2017}), \bibinfo{numpages}{31}~pages.
\newblock


\bibitem[Alvaro et~al\mbox{.}(2011)]%
        {alvaro2011consistency}
\bibfield{author}{\bibinfo{person}{Peter Alvaro}, \bibinfo{person}{Neil
  Conway}, \bibinfo{person}{Joseph~M Hellerstein}, {and}
  \bibinfo{person}{William~R Marczak}.} \bibinfo{year}{2011}\natexlab{}.
\newblock \showarticletitle{Consistency Analysis in Bloom: a CALM and Collected
  Approach.}. In \bibinfo{booktitle}{\emph{CIDR}}. \bibinfo{pages}{249--260}.
\newblock


\bibitem[Amiri et~al\mbox{.}(2019)]%
        {amiri2019caper}
\bibfield{author}{\bibinfo{person}{Mohammad~Javad Amiri},
  \bibinfo{person}{Divyakant Agrawal}, {and} \bibinfo{person}{Amr~El Abbadi}.}
  \bibinfo{year}{2019}\natexlab{}.
\newblock \showarticletitle{Caper: a cross-application permissioned
  blockchain}.
\newblock \bibinfo{journal}{\emph{Proceedings of the VLDB Endowment}}
  \bibinfo{volume}{12}, \bibinfo{number}{11} (\bibinfo{year}{2019}),
  \bibinfo{pages}{1385--1398}.
\newblock


\bibitem[Amiri et~al\mbox{.}(2021)]%
        {amiri2021sharper}
\bibfield{author}{\bibinfo{person}{Mohammad~Javad Amiri},
  \bibinfo{person}{Divyakant Agrawal}, {and} \bibinfo{person}{Amr El~Abbadi}.}
  \bibinfo{year}{2021}\natexlab{}.
\newblock \showarticletitle{Sharper: Sharding permissioned blockchains over
  network clusters}. In \bibinfo{booktitle}{\emph{Proceedings of the 2021
  International Conference on Management of Data}}. \bibinfo{pages}{76--88}.
\newblock


\bibitem[Androulaki et~al\mbox{.}(2018)]%
        {androulaki2018hyperledger}
\bibfield{author}{\bibinfo{person}{Elli Androulaki}, \bibinfo{person}{Artem
  Barger}, \bibinfo{person}{Vita Bortnikov}, \bibinfo{person}{Christian
  Cachin}, \bibinfo{person}{Konstantinos Christidis}, \bibinfo{person}{Angelo
  De~Caro}, \bibinfo{person}{David Enyeart}, \bibinfo{person}{Christopher
  Ferris}, \bibinfo{person}{Gennady Laventman}, \bibinfo{person}{Yacov
  Manevich}, {et~al\mbox{.}}} \bibinfo{year}{2018}\natexlab{}.
\newblock \showarticletitle{Hyperledger fabric: a distributed operating system
  for permissioned blockchains}. In \bibinfo{booktitle}{\emph{Proceedings of
  the thirteenth EuroSys conference}}. \bibinfo{pages}{1--15}.
\newblock


\bibitem[Avni et~al\mbox{.}(2020)]%
        {mot}
\bibfield{author}{\bibinfo{person}{H. Avni}, \bibinfo{person}{A. Aliev},
  \bibinfo{person}{O. Amor}, \bibinfo{person}{A. Avitzur}, \bibinfo{person}{I.
  Bronshtein}, \bibinfo{person}{E. Ginot}, \bibinfo{person}{S. Goikhman},
  \bibinfo{person}{E. Levy}, \bibinfo{person}{Lu Levy, I.},
  \bibinfo{person}{F.}, {and} \bibinfo{person}{L. Mishali}.}
  \bibinfo{year}{2020}\natexlab{}.
\newblock \showarticletitle{Industrial-Strength OLTP Using Main Memory and Many
  Cores}.
\newblock \bibinfo{journal}{\emph{Proceedings of the VLDB Endowment}}
  \bibinfo{volume}{13}, \bibinfo{number}{12} (\bibinfo{year}{2020}),
  \bibinfo{pages}{3099--3111}.
\newblock


\bibitem[Bailis et~al\mbox{.}(2014)]%
        {10.14778/2735508.2735509}
\bibfield{author}{\bibinfo{person}{Peter Bailis}, \bibinfo{person}{Alan
  Fekete}, \bibinfo{person}{Michael~J. Franklin}, \bibinfo{person}{Ali Ghodsi},
  \bibinfo{person}{Joseph~M. Hellerstein}, {and} \bibinfo{person}{Ion Stoica}.}
  \bibinfo{year}{2014}\natexlab{}.
\newblock \showarticletitle{Coordination Avoidance in Database Systems}.
\newblock \bibinfo{journal}{\emph{Proc. VLDB Endow.}} \bibinfo{volume}{8},
  \bibinfo{number}{3} (\bibinfo{year}{2014}), \bibinfo{pages}{185–196}.
\newblock


\bibitem[Bailis(2015)]%
        {bailis2015coordination}
\bibfield{author}{\bibinfo{person}{Peter~David Bailis}.}
  \bibinfo{year}{2015}\natexlab{}.
\newblock \bibinfo{booktitle}{\emph{Coordination avoidance in distributed
  databases}}.
\newblock \bibinfo{publisher}{University of California, Berkeley}.
\newblock


\bibitem[Cahill et~al\mbox{.}(2009)]%
        {cahill2009serializable}
\bibfield{author}{\bibinfo{person}{Michael~J Cahill}, \bibinfo{person}{Uwe
  R{\"o}hm}, {and} \bibinfo{person}{Alan~D Fekete}.}
  \bibinfo{year}{2009}\natexlab{}.
\newblock \showarticletitle{Serializable isolation for snapshot databases}.
\newblock \bibinfo{journal}{\emph{ACM Transactions on Database Systems (TODS)}}
  \bibinfo{volume}{34}, \bibinfo{number}{4} (\bibinfo{year}{2009}),
  \bibinfo{pages}{1--42}.
\newblock


\bibitem[Chairunnanda et~al\mbox{.}(2014)]%
        {chairunnanda2014confluxdb}
\bibfield{author}{\bibinfo{person}{Prima Chairunnanda},
  \bibinfo{person}{Khuzaima Daudjee}, {and} \bibinfo{person}{M~Tamer
  {\"O}zsu}.} \bibinfo{year}{2014}\natexlab{}.
\newblock \showarticletitle{ConfluxDB: Multi-master replication for partitioned
  snapshot isolation databases}.
\newblock \bibinfo{journal}{\emph{Proceedings of the VLDB Endowment}}
  \bibinfo{volume}{7}, \bibinfo{number}{11} (\bibinfo{year}{2014}),
  \bibinfo{pages}{947--958}.
\newblock


\bibitem[Conway et~al\mbox{.}(2012)]%
        {Conway:2012:LLD:2391229.2391230}
\bibfield{author}{\bibinfo{person}{Neil Conway}, \bibinfo{person}{William~R.
  Marczak}, \bibinfo{person}{Peter Alvaro}, \bibinfo{person}{Joseph~M.
  Hellerstein}, {and} \bibinfo{person}{David Maier}.}
  \bibinfo{year}{2012}\natexlab{}.
\newblock \showarticletitle{Logic and Lattices for Distributed Programming}. In
  \bibinfo{booktitle}{\emph{Proceedings of the Symposium on Cloud Computing
  (SoCC '12)}}. \bibinfo{pages}{1:1--1:14}.
\newblock


\bibitem[Cooper et~al\mbox{.}(2010)]%
        {10.1145/1807128.1807152}
\bibfield{author}{\bibinfo{person}{Brian~F. Cooper}, \bibinfo{person}{Adam
  Silberstein}, \bibinfo{person}{Erwin Tam}, \bibinfo{person}{Raghu
  Ramakrishnan}, {and} \bibinfo{person}{Russell Sears}.}
  \bibinfo{year}{2010}\natexlab{}.
\newblock \showarticletitle{Benchmarking Cloud Serving Systems with YCSB}. In
  \bibinfo{booktitle}{\emph{Proceedings of the 1st ACM Symposium on Cloud
  Computing}} \emph{(\bibinfo{series}{SoCC '10})}. \bibinfo{pages}{143–154}.
\newblock


\bibitem[Corbett et~al\mbox{.}(2013)]%
        {corbett2013spanner}
\bibfield{author}{\bibinfo{person}{James~C Corbett}, \bibinfo{person}{Jeffrey
  Dean}, \bibinfo{person}{Michael Epstein}, \bibinfo{person}{Andrew Fikes},
  \bibinfo{person}{Christopher Frost}, \bibinfo{person}{Jeffrey~John Furman},
  \bibinfo{person}{Sanjay Ghemawat}, \bibinfo{person}{Andrey Gubarev},
  \bibinfo{person}{Christopher Heiser}, \bibinfo{person}{Peter Hochschild},
  {et~al\mbox{.}}} \bibinfo{year}{2013}\natexlab{}.
\newblock \showarticletitle{Spanner: Google’s globally distributed database}.
\newblock \bibinfo{journal}{\emph{ACM Transactions on Computer Systems (TOCS)}}
  \bibinfo{volume}{31}, \bibinfo{number}{3} (\bibinfo{year}{2013}),
  \bibinfo{pages}{1--22}.
\newblock


\bibitem[Dang et~al\mbox{.}(2019)]%
        {dang2019towards}
\bibfield{author}{\bibinfo{person}{Hung Dang}, \bibinfo{person}{Tien Tuan~Anh
  Dinh}, \bibinfo{person}{Dumitrel Loghin}, \bibinfo{person}{Ee-Chien Chang},
  \bibinfo{person}{Qian Lin}, {and} \bibinfo{person}{Beng~Chin Ooi}.}
  \bibinfo{year}{2019}\natexlab{}.
\newblock \showarticletitle{Towards scaling blockchain systems via sharding}.
  In \bibinfo{booktitle}{\emph{Proceedings of the 2019 international conference
  on management of data}}. \bibinfo{pages}{123--140}.
\newblock


\bibitem[DeCandia et~al\mbox{.}(2007)]%
        {decandia2007dynamo}
\bibfield{author}{\bibinfo{person}{Giuseppe DeCandia}, \bibinfo{person}{Deniz
  Hastorun}, \bibinfo{person}{Madan Jampani}, \bibinfo{person}{Gunavardhan
  Kakulapati}, \bibinfo{person}{Avinash Lakshman}, \bibinfo{person}{Alex
  Pilchin}, \bibinfo{person}{Swaminathan Sivasubramanian},
  \bibinfo{person}{Peter Vosshall}, {and} \bibinfo{person}{Werner Vogels}.}
  \bibinfo{year}{2007}\natexlab{}.
\newblock \showarticletitle{Dynamo: Amazon's highly available key-value store}.
\newblock \bibinfo{journal}{\emph{ACM SIGOPS operating systems review}}
  \bibinfo{volume}{41}, \bibinfo{number}{6} (\bibinfo{year}{2007}),
  \bibinfo{pages}{205--220}.
\newblock


\bibitem[Elnikety et~al\mbox{.}(2006)]%
        {10.1145/1217935.1217947}
\bibfield{author}{\bibinfo{person}{Sameh Elnikety}, \bibinfo{person}{Steven
  Dropsho}, {and} \bibinfo{person}{Fernando Pedone}.}
  \bibinfo{year}{2006}\natexlab{}.
\newblock \showarticletitle{Tashkent: Uniting Durability with Transaction
  Ordering for High-Performance Scalable Database Replication}. In
  \bibinfo{booktitle}{\emph{Proceedings of the 1st ACM SIGOPS/EuroSys European
  Conference on Computer Systems 2006 (EuroSys '06)}}
  \emph{(\bibinfo{series}{EuroSys '06})}. \bibinfo{pages}{117–130}.
\newblock


\bibitem[Faleiro et~al\mbox{.}(2017)]%
        {faleiro2017high}
\bibfield{author}{\bibinfo{person}{Jose~M Faleiro}, \bibinfo{person}{Daniel~J
  Abadi}, {and} \bibinfo{person}{Joseph~M Hellerstein}.}
  \bibinfo{year}{2017}\natexlab{}.
\newblock \showarticletitle{High performance transactions via early write
  visibility}.
\newblock \bibinfo{journal}{\emph{Proceedings of the VLDB Endowment}}
  \bibinfo{volume}{10}, \bibinfo{number}{5} (\bibinfo{year}{2017}).
\newblock


\bibitem[group(2022)]%
        {oceanbase}
\bibfield{author}{\bibinfo{person}{Ant group}.}
  \bibinfo{year}{2022}\natexlab{}.
\newblock \bibinfo{title}{OceanBase}.
\newblock
\newblock
\urldef\tempurl%
\url{https://open.oceanbase.com/}
\showURL{%
\tempurl}


\bibitem[Gupta et~al\mbox{.}(2020)]%
        {gupta2020resilientdb}
\bibfield{author}{\bibinfo{person}{Suyash Gupta}, \bibinfo{person}{Sajjad
  Rahnama}, \bibinfo{person}{Jelle Hellings}, {and} \bibinfo{person}{Mohammad
  Sadoghi}.} \bibinfo{year}{2020}\natexlab{}.
\newblock \showarticletitle{Resilientdb: Global scale resilient blockchain
  fabric}.
\newblock \bibinfo{journal}{\emph{arXiv preprint arXiv:2002.00160}}
  (\bibinfo{year}{2020}).
\newblock


\bibitem[Harding et~al\mbox{.}(2017)]%
        {harding2017evaluation}
\bibfield{author}{\bibinfo{person}{Rachael Harding}, \bibinfo{person}{Dana
  Van~Aken}, \bibinfo{person}{Andrew Pavlo}, {and} \bibinfo{person}{Michael
  Stonebraker}.} \bibinfo{year}{2017}\natexlab{}.
\newblock \showarticletitle{An evaluation of distributed concurrency control}.
\newblock \bibinfo{journal}{\emph{Proceedings of the VLDB Endowment}}
  \bibinfo{volume}{10}, \bibinfo{number}{5} (\bibinfo{year}{2017}),
  \bibinfo{pages}{553--564}.
\newblock


\bibitem[Hellings and Sadoghi(2021)]%
        {hellings2021byshard}
\bibfield{author}{\bibinfo{person}{Jelle Hellings} {and}
  \bibinfo{person}{Mohammad Sadoghi}.} \bibinfo{year}{2021}\natexlab{}.
\newblock \showarticletitle{Byshard: Sharding in a byzantine environment}.
\newblock \bibinfo{journal}{\emph{Proceedings of the VLDB Endowment}}
  \bibinfo{volume}{14}, \bibinfo{number}{11} (\bibinfo{year}{2021}),
  \bibinfo{pages}{2230--2243}.
\newblock


\bibitem[Lakshman and Malik(2010)]%
        {lakshman2010cassandra}
\bibfield{author}{\bibinfo{person}{Avinash Lakshman} {and}
  \bibinfo{person}{Prashant Malik}.} \bibinfo{year}{2010}\natexlab{}.
\newblock \showarticletitle{Cassandra: a decentralized structured storage
  system}.
\newblock \bibinfo{journal}{\emph{ACM SIGOPS Operating Systems Review}}
  \bibinfo{volume}{44}, \bibinfo{number}{2} (\bibinfo{year}{2010}),
  \bibinfo{pages}{35--40}.
\newblock


\bibitem[Lu et~al\mbox{.}(2020)]%
        {lu2020aria}
\bibfield{author}{\bibinfo{person}{Yi Lu}, \bibinfo{person}{Xiangyao Yu},
  \bibinfo{person}{Lei Cao}, {and} \bibinfo{person}{Samuel Madden}.}
  \bibinfo{year}{2020}\natexlab{}.
\newblock \showarticletitle{Aria: a fast and practical deterministic OLTP
  database}.
\newblock \bibinfo{journal}{\emph{Proceedings of the VLDB Endowment}}
  \bibinfo{volume}{13}, \bibinfo{number}{12} (\bibinfo{year}{2020}),
  \bibinfo{pages}{2047--2060}.
\newblock


\bibitem[Lu et~al\mbox{.}(2021)]%
        {10.14778/3446095.3446098}
\bibfield{author}{\bibinfo{person}{Yi Lu}, \bibinfo{person}{Xiangyao Yu},
  \bibinfo{person}{Lei Cao}, {and} \bibinfo{person}{Samuel Madden}.}
  \bibinfo{year}{2021}\natexlab{}.
\newblock \showarticletitle{Epoch-Based Commit and Replication in Distributed
  OLTP Databases}.
\newblock \bibinfo{journal}{\emph{Proc. VLDB Endow.}} \bibinfo{volume}{14},
  \bibinfo{number}{5} (\bibinfo{year}{2021}), \bibinfo{pages}{743–756}.
\newblock


\bibitem[Lu et~al\mbox{.}(2019)]%
        {10.14778/3342263.3342270}
\bibfield{author}{\bibinfo{person}{Yi Lu}, \bibinfo{person}{Xiangyao Yu}, {and}
  \bibinfo{person}{Samuel Madden}.} \bibinfo{year}{2019}\natexlab{}.
\newblock \showarticletitle{STAR: Scaling Transactions through Asymmetric
  Replication}.
\newblock \bibinfo{journal}{\emph{Proc. VLDB Endow.}} \bibinfo{volume}{12},
  \bibinfo{number}{11} (\bibinfo{year}{2019}), \bibinfo{pages}{1316–1329}.
\newblock


\bibitem[Nakamoto(2008)]%
        {nakamoto2008bitcoin}
\bibfield{author}{\bibinfo{person}{Satoshi Nakamoto}.}
  \bibinfo{year}{2008}\natexlab{}.
\newblock \showarticletitle{Bitcoin: A peer-to-peer electronic cash system}.
\newblock \bibinfo{journal}{\emph{Decentralized Business Review}}
  (\bibinfo{year}{2008}), \bibinfo{pages}{21260}.
\newblock


\bibitem[Pincap(2022)]%
        {tidb}
\bibfield{author}{\bibinfo{person}{Pincap}.} \bibinfo{year}{2022}\natexlab{}.
\newblock \bibinfo{title}{TiDB}.
\newblock
\newblock
\urldef\tempurl%
\url{https://pingcap.com/products/tidb}
\showURL{%
\tempurl}


\bibitem[Pregui{\c{c}}a(2018)]%
        {preguicca2018conflict}
\bibfield{author}{\bibinfo{person}{Nuno Pregui{\c{c}}a}.}
  \bibinfo{year}{2018}\natexlab{}.
\newblock \showarticletitle{Conflict-free replicated data types: An overview}.
\newblock \bibinfo{journal}{\emph{arXiv preprint arXiv:1806.10254}}
  (\bibinfo{year}{2018}).
\newblock


\bibitem[Qadah et~al\mbox{.}(2020)]%
        {qadah2020q}
\bibfield{author}{\bibinfo{person}{Thamir Qadah}, \bibinfo{person}{Suyash
  Gupta}, {and} \bibinfo{person}{Mohammad Sadoghi}.}
  \bibinfo{year}{2020}\natexlab{}.
\newblock \showarticletitle{Q-Store: Distributed, Multi-partition Transactions
  via Queue-oriented Execution and Communication}. In
  \bibinfo{booktitle}{\emph{Proceedings of the 23rd International Conference on
  Extending Database Technology (EDBT)}}. \bibinfo{pages}{73--84}.
\newblock


\bibitem[Qadah and Sadoghi(2018)]%
        {qadah2018quecc}
\bibfield{author}{\bibinfo{person}{Thamir~M Qadah} {and}
  \bibinfo{person}{Mohammad Sadoghi}.} \bibinfo{year}{2018}\natexlab{}.
\newblock \showarticletitle{Quecc: A queue-oriented, control-free concurrency
  architecture}. In \bibinfo{booktitle}{\emph{Proceedings of the 19th
  International Middleware Conference}}. \bibinfo{pages}{13--25}.
\newblock


\bibitem[Rae et~al\mbox{.}(2013)]%
        {10.14778/2536222.2536230}
\bibfield{author}{\bibinfo{person}{Ian Rae}, \bibinfo{person}{Eric Rollins},
  \bibinfo{person}{Jeff Shute}, \bibinfo{person}{Sukhdeep Sodhi}, {and}
  \bibinfo{person}{Radek Vingralek}.} \bibinfo{year}{2013}\natexlab{}.
\newblock \showarticletitle{Online, Asynchronous Schema Change in F1}.
\newblock \bibinfo{journal}{\emph{Proc. VLDB Endow.}} \bibinfo{volume}{6},
  \bibinfo{number}{11} (\bibinfo{date}{aug} \bibinfo{year}{2013}),
  \bibinfo{pages}{1045–1056}.
\newblock


\bibitem[Rahnama et~al\mbox{.}(2021)]%
        {rahnama2021ringbft}
\bibfield{author}{\bibinfo{person}{Sajjad Rahnama}, \bibinfo{person}{Suyash
  Gupta}, \bibinfo{person}{Rohan Sogani}, \bibinfo{person}{Dhruv Krishnan},
  {and} \bibinfo{person}{Mohammad Sadoghi}.} \bibinfo{year}{2021}\natexlab{}.
\newblock \showarticletitle{RingBFT: Resilient Consensus over Sharded Ring
  Topology}.
\newblock \bibinfo{journal}{\emph{arXiv preprint arXiv:2107.13047}}
  (\bibinfo{year}{2021}).
\newblock


\bibitem[Ren et~al\mbox{.}(2019)]%
        {10.14778/3342263.3342647}
\bibfield{author}{\bibinfo{person}{Kun Ren}, \bibinfo{person}{Dennis Li}, {and}
  \bibinfo{person}{Daniel~J. Abadi}.} \bibinfo{year}{2019}\natexlab{}.
\newblock \showarticletitle{SLOG: Serializable, Low-Latency, Geo-Replicated
  Transactions}.
\newblock \bibinfo{journal}{\emph{Proc. VLDB Endow.}} \bibinfo{volume}{12},
  \bibinfo{number}{11} (\bibinfo{date}{jul} \bibinfo{year}{2019}),
  \bibinfo{pages}{1747–1761}.
\newblock


\bibitem[Ren et~al\mbox{.}(2014)]%
        {ren2014evaluation}
\bibfield{author}{\bibinfo{person}{Kun Ren}, \bibinfo{person}{Alexander
  Thomson}, {and} \bibinfo{person}{Daniel~J Abadi}.}
  \bibinfo{year}{2014}\natexlab{}.
\newblock \showarticletitle{An evaluation of the advantages and disadvantages
  of deterministic database systems}.
\newblock \bibinfo{journal}{\emph{Proceedings of the VLDB Endowment}}
  \bibinfo{volume}{7}, \bibinfo{number}{10} (\bibinfo{year}{2014}),
  \bibinfo{pages}{821--832}.
\newblock


\bibitem[Shapiro et~al\mbox{.}(2011a)]%
        {shapiro2011comprehensive}
\bibfield{author}{\bibinfo{person}{Marc Shapiro}, \bibinfo{person}{Nuno
  Pregui{\c{c}}a}, \bibinfo{person}{Carlos Baquero}, {and}
  \bibinfo{person}{Marek Zawirski}.} \bibinfo{year}{2011}\natexlab{a}.
\newblock \emph{\bibinfo{title}{A comprehensive study of convergent and
  commutative replicated data types}}.
\newblock \bibinfo{thesistype}{Ph.\,D. Dissertation}.
  \bibinfo{school}{Inria--Centre Paris-Rocquencourt; INRIA}.
\newblock


\bibitem[Shapiro et~al\mbox{.}(2011b)]%
        {shapiro2011conflict}
\bibfield{author}{\bibinfo{person}{Marc Shapiro}, \bibinfo{person}{Nuno
  Pregui{\c{c}}a}, \bibinfo{person}{Carlos Baquero}, {and}
  \bibinfo{person}{Marek Zawirski}.} \bibinfo{year}{2011}\natexlab{b}.
\newblock \showarticletitle{Conflict-free replicated data types}. In
  \bibinfo{booktitle}{\emph{Symposium on Self-Stabilizing Systems}}.
  \bibinfo{pages}{386--400}.
\newblock


\bibitem[Shapiro et~al\mbox{.}(2011c)]%
        {Shapiro:2011:CRD:2050613.2050642}
\bibfield{author}{\bibinfo{person}{Marc Shapiro}, \bibinfo{person}{Nuno
  Pregui\c{c}a}, \bibinfo{person}{Carlos Baquero}, {and} \bibinfo{person}{Marek
  Zawirski}.} \bibinfo{year}{2011}\natexlab{c}.
\newblock \showarticletitle{Conflict-free Replicated Data Types}. In
  \bibinfo{booktitle}{\emph{Proceedings of the Symposium on Self-stabilizing
  Systems (SSS '11)}}. \bibinfo{pages}{386--400}.
\newblock


\bibitem[Stathakopoulou et~al\mbox{.}(2022)]%
        {stathakopoulou2022state}
\bibfield{author}{\bibinfo{person}{Chrysoula Stathakopoulou},
  \bibinfo{person}{Matej Pavlovic}, {and} \bibinfo{person}{Marko Vukoli{\'c}}.}
  \bibinfo{year}{2022}\natexlab{}.
\newblock \showarticletitle{State machine replication scalability made simple}.
  In \bibinfo{booktitle}{\emph{Proceedings of the Seventeenth European
  Conference on Computer Systems}}. \bibinfo{pages}{17--33}.
\newblock


\bibitem[Taft et~al\mbox{.}(2020)]%
        {taft2020cockroachdb}
\bibfield{author}{\bibinfo{person}{Rebecca Taft}, \bibinfo{person}{Irfan
  Sharif}, \bibinfo{person}{Andrei Matei}, \bibinfo{person}{Nathan
  VanBenschoten}, \bibinfo{person}{Jordan Lewis}, \bibinfo{person}{Tobias
  Grieger}, \bibinfo{person}{Kai Niemi}, \bibinfo{person}{Andy Woods},
  \bibinfo{person}{Anne Birzin}, \bibinfo{person}{Raphael Poss},
  {et~al\mbox{.}}} \bibinfo{year}{2020}\natexlab{}.
\newblock \showarticletitle{Cockroachdb: The resilient geo-distributed sql
  database}. In \bibinfo{booktitle}{\emph{Proceedings of the 2020 ACM SIGMOD
  International Conference on Management of Data}}.
  \bibinfo{pages}{1493--1509}.
\newblock


\bibitem[Thomson and Abadi(2010)]%
        {thomson2010case}
\bibfield{author}{\bibinfo{person}{Alexander Thomson} {and}
  \bibinfo{person}{Daniel~J Abadi}.} \bibinfo{year}{2010}\natexlab{}.
\newblock \showarticletitle{The case for determinism in database systems}.
\newblock \bibinfo{journal}{\emph{Proceedings of the VLDB Endowment}}
  \bibinfo{volume}{3}, \bibinfo{number}{1-2} (\bibinfo{year}{2010}),
  \bibinfo{pages}{70--80}.
\newblock


\bibitem[Thomson and Abadi(2015)]%
        {thomson2015calvinfs}
\bibfield{author}{\bibinfo{person}{Alexander Thomson} {and}
  \bibinfo{person}{Daniel~J Abadi}.} \bibinfo{year}{2015}\natexlab{}.
\newblock \showarticletitle{CalvinFS: Consistent WAN Replication and Scalable
  Metadata Management for Distributed File Systems}. In
  \bibinfo{booktitle}{\emph{Proceedings of the 13th USENIX Conference on File
  and Storage Technologies (FAST 15)}}. \bibinfo{pages}{1--14}.
\newblock


\bibitem[Thomson et~al\mbox{.}(2012)]%
        {thomson2012calvin}
\bibfield{author}{\bibinfo{person}{Alexander Thomson},
  \bibinfo{person}{Thaddeus Diamond}, \bibinfo{person}{Shu-Chun Weng},
  \bibinfo{person}{Kun Ren}, \bibinfo{person}{Philip Shao}, {and}
  \bibinfo{person}{Daniel~J Abadi}.} \bibinfo{year}{2012}\natexlab{}.
\newblock \showarticletitle{Calvin: fast distributed transactions for
  partitioned database systems}. In \bibinfo{booktitle}{\emph{Proceedings of
  the 2012 ACM SIGMOD International Conference on Management of Data}}.
  \bibinfo{pages}{1--12}.
\newblock


\bibitem[Verbitski et~al\mbox{.}(2017)]%
        {10.1145/3035918.3056101}
\bibfield{author}{\bibinfo{person}{Alexandre Verbitski},
  \bibinfo{person}{Anurag Gupta}, \bibinfo{person}{Debanjan Saha},
  \bibinfo{person}{Murali Brahmadesam}, \bibinfo{person}{Kamal Gupta},
  \bibinfo{person}{Raman Mittal}, \bibinfo{person}{Sailesh Krishnamurthy},
  \bibinfo{person}{Sandor Maurice}, \bibinfo{person}{Tengiz Kharatishvili},
  {and} \bibinfo{person}{Xiaofeng Bao}.} \bibinfo{year}{2017}\natexlab{}.
\newblock \showarticletitle{Amazon Aurora: Design Considerations for High
  Throughput Cloud-Native Relational Databases}. In
  \bibinfo{booktitle}{\emph{Proceedings of the 2017 ACM International
  Conference on Management of Data}} \emph{(\bibinfo{series}{SIGMOD '17})}.
  \bibinfo{pages}{1041–1052}.
\newblock


\bibitem[Wu et~al\mbox{.}(2019a)]%
        {wu2019anna}
\bibfield{author}{\bibinfo{person}{Chenggang Wu}, \bibinfo{person}{Jose~M
  Faleiro}, \bibinfo{person}{Yihan Lin}, {and} \bibinfo{person}{Joseph~M
  Hellerstein}.} \bibinfo{year}{2019}\natexlab{a}.
\newblock \showarticletitle{Anna: A kvs for any scale}.
\newblock \bibinfo{journal}{\emph{IEEE Transactions on Knowledge and Data
  Engineering}} \bibinfo{volume}{33}, \bibinfo{number}{2}
  (\bibinfo{year}{2019}), \bibinfo{pages}{344--358}.
\newblock


\bibitem[Wu et~al\mbox{.}(2019b)]%
        {wu2019autoscaling}
\bibfield{author}{\bibinfo{person}{Chenggang Wu}, \bibinfo{person}{Vikram
  Sreekanti}, {and} \bibinfo{person}{Joseph~M Hellerstein}.}
  \bibinfo{year}{2019}\natexlab{b}.
\newblock \showarticletitle{Autoscaling tiered cloud storage in Anna}.
\newblock \bibinfo{journal}{\emph{Proceedings of the VLDB Endowment}}
  \bibinfo{volume}{12}, \bibinfo{number}{6} (\bibinfo{year}{2019}),
  \bibinfo{pages}{624--638}.
\newblock


\bibitem[Yao et~al\mbox{.}(2016)]%
        {yao2016exploiting}
\bibfield{author}{\bibinfo{person}{Chang Yao}, \bibinfo{person}{Divyakant
  Agrawal}, \bibinfo{person}{Gang Chen}, \bibinfo{person}{Qian Lin},
  \bibinfo{person}{Beng~Chin Ooi}, \bibinfo{person}{Weng-Fai Wong}, {and}
  \bibinfo{person}{Meihui Zhang}.} \bibinfo{year}{2016}\natexlab{}.
\newblock \showarticletitle{Exploiting single-threaded model in multi-core
  in-memory systems}.
\newblock \bibinfo{journal}{\emph{IEEE Transactions on Knowledge and Data
  Engineering}} \bibinfo{volume}{28}, \bibinfo{number}{10}
  (\bibinfo{year}{2016}), \bibinfo{pages}{2635--2650}.
\newblock


\end{thebibliography}
